\documentclass[11pt]{article}
\usepackage{epsfig}
\usepackage{amsfonts}
\usepackage{amsmath}
\usepackage{amssymb}
\usepackage{bbm,bm}
\usepackage{graphicx}
\usepackage{slashed}
\allowdisplaybreaks[1]

\usepackage{tikz}
\newdimen\nodeDist
\usepackage{datetime}
\usepackage{xcolor}
\usepackage{hyperref}
\usepackage[numbers,sort&compress]{natbib}
\input pix.sty

\renewcommand{\eq}{eq.~}
\renewcommand{\eqs}{eqs.~}

\renewcommand{\fig}{fig.~}
\renewcommand{\figs}{figs.~}

\newcommand{\mD}{m_\rmii{D}}

\newcommand{\gammaE}{\gamma_\rmii{E}}

\newcommand{\rmO}{{\mathcal{O}}}

\def\lsi{\raise0.3ex\hbox{$<$\kern-0.75em\raise-1.1ex\hbox{$\sim$}}}
\def\gsi{\raise0.3ex\hbox{$>$\kern-0.75em\raise-1.1ex\hbox{$\sim$}}}

\newcommand{\sign}{\mathop{\mbox{sign}}}
\newcommand{\nF}{n_\rmii{F}}
\newcommand{\nB}{n_\rmii{B}}
\renewcommand{\P}{\mathcal{P}}
\newcommand{\K}{\mathcal{K}}
\newcommand{\Q}{\mathcal{Q}}
\newcommand{\rmii}[1]{{\mbox{\tiny\rm{#1}}}}

\newcommand{\Tint}[1]{{\hbox{$\sum$}\!\!\!\!\!\!\!\int\,}_{\!\!\!\!\raise-0.9ex\hbox{$\scriptstyle{#1}$}}}
\newcommand{\Tinti}[1]{{{\Sigma}\!\!\!\!\raise0.3ex\hbox{$\int$}_\rmii{${#1}$}}}

\newcommand{\bi}{\begin{itemize}}
\newcommand{\ei}{\end{itemize}}
\newcommand{\hide}[1]{ }

\def\mpl{m_\rmi{Pl}}
\def\acrs{\textsc{AutoTherm}}
\def\mDi{m_{\rmii{D}i}}
\def\ncol{N_i}
\def\nscal{N_{s\,i}}
\def\nferm{N_{f\,i}}
\def\twotwo{2\leftrightarrow 2}
\makeatletter \@addtoreset{equation}{section} \makeatother
\renewcommand{\theequation}{\arabic{section}.\arabic{equation}}
\makeatletter
\renewcommand\section{\@startsection{section}{1}{\z@}%
  {-5.5ex \@plus -1ex \@minus -.2ex}
  {2.3ex \@plus.2ex}%
  {\normalfont\large\bfseries}}
\renewcommand\subsection{\@startsection{subsection}{2}{\z@}%
  {-3.25ex\@plus -1ex \@minus -.2ex}%
  {1.5ex \@plus .2ex}%
  {\normalfont\normalsize\bfseries}}
\renewcommand\thesection{\@arabic\c@section}
\renewcommand\thesubsection{\thesection.\@arabic\c@subsection}
\renewcommand{\@seccntformat}[1]{%
  \csname the#1\endcsname.\hspace{1.0em}}
\makeatother
\begin{document}

\flushbottom

\begin{titlepage}

\begin{flushright}
September~2026
\end{flushright}
\begin{centering}

\vfill

{\Large{\bf
Thermal gravitino and axino rate from AutoTherm
}}

\vspace{0.8cm}
Killian~Bouzoud$^{\rm a}$,
Jacopo~Ghiglieri$^{\rm a}$ 
and Greg~Jackson$^{\rm a}$ 

\vspace{0.8cm}

$^{\rm a}$%
{\em
SUBATECH, Nantes Universit\'e, IMT Atlantique, IN2P3/CNRS,\\
4 rue Alfred Kastler, La Chantrerie BP 20722, 44307 Nantes, France\\
}

\vspace*{0.8cm}

\mbox{\bf Abstract}

\end{centering}

\vspace*{0.3cm}

\noindent
If ultrarelativistic supersymmetric particles are 
thermally produced in 
the early universe they would influence big-bang nucleosynthesis, 
dark matter and 
reheating constraints. 
While the basic framework for thermal gravitino and axino production 
is well established,
the calculation is not completely settled. 
In this work, we revisit the problem using \acrs{}, 
a new tool which automates the computation of thermal rates from 
first-principles Thermal Field Theory,  with control over
the inherent theory uncertainty in 
Hard Thermal Loop resummation schemes.
We demonstrate that the strict leading-order (LO) scheme, 
while unambiguous for hard momenta, 
can yield unphysical negative rates when extrapolated to soft momenta. 
To address this, we introduce a tuned scheme that 
ensures positivity and agreement with strict LO at high momenta, 
while avoiding gauge-dependent pathologies. 
We compare our results with existing parametrizations in 
the literature, identifying discrepancies and providing new, 
reliable fits for the gauge and Yukawa contributions to 
the gravitino and axino production rates. 
The choice of resummation scheme gives a theory 
uncertainty 
by a factor of $\sim 1.5$...$3$ depending on the temperature, 
which has been underappreciated until now. 
This residual spread should be taken into account in 
precision cosmological analyses of gravitino/axino abundancies.

\vfill
\end{titlepage}

\tableofcontents

%
\section{Introduction}

The gravitino arises naturally in supersymmetry (SUSY) once it is extended locally to supergravity (SUGRA) 
\cite{Deser:1976eh,Freedman:1976xh}. 
As the supersymmetric partner of the graviton, it takes the form of a spin 3/2 fermion. Unlike the graviton,
however, it acquires a mass $m_{3/2}$ through SUSY breaking, thus representing a possible 
dark matter (DM) candidate \cite{Ellis:1984eq,Moroi:1993mb,Giudice:1999am} if it is the lightest
supersymmetric particle.

Just like the graviton, the gravitino couples rather universally to radiation and matter through Planck-mass ($\mpl\equiv 1/\sqrt{G}$) 
suppressed dimension-five operators. This has two consequences; first, current DM searches would be insensitive
to it. Second, its thermal production rate in the 
Early Universe 
when supersymmetry is restored, $T \gg m_{3/2}\,$, 
would scale on dimensional grounds
like $T^3/\mpl^2$. 
As the Hubble rate in the radiation epoch scales like $T^2/\mpl$,
thermal production would happen mostly at and shortly after reheating, in a classic case of 
\emph{ultraviolet freeze in}.

Hence, at fixed $m_{3/2}$, higher reheating temperatures $T_\mathrm{RH}$ will result 
in larger DM abundances, leading to the so-called \emph{gravitino overproduction bound} 
\cite{Ellis:1984eq,Moroi:1993mb,Giudice:1999am} from overclosure. In scenarios
where the gravitino is unstable, one speaks generically of a \emph{gravitino problem}
\cite{Weinberg:1982zq,Ellis:1983ew}
whenever the consequences of its early production and later decays would spoil established
elements of the cosmological history of the early universe, such as Big-Bang Nucleosynthesis.

This has motivated 
increasingly sophisticated calculations of the main ingredient for the bound and more generally 
for the gravitino abundance, the 
\emph{gravitino thermal production rate} for gravitino momenta $k$ satisfying $k\gtrsim T\gg m_{3/2}$.
\cite{Ellis:1984eq,Moroi:1993mb} introduced a fixed gauge-boson thermal mass to regulate the 
infrared divergence arising from Coulomb-like gauge-boson mediated processes
such as gaugino gaugino going into gaugino gravitino. 
This was later turned into a complete, \emph{strict} leading-order (LO) calculation
in \cite{Bolz:2000fu,Pradler:2006qh,Pradler:2006hh,Pradler:2006tpx}, using the method 
of \cite{Braaten:1991dd} to consistently account for the 
emergence of \emph{collective effects} at the \emph{soft scale} $g_i T$, with $g_i$
the gauge couplings, through Hard Thermal Loop (HTL) resummation~\cite{Braaten:1989mz}. 
The \emph{dynamical screening}, also known as Landau damping, 
originating from these effects screens the would-be naked divergence making the rate finite.

The emergence of this second, soft scale besides the 
\emph{hard scale}  $k\gtrsim T$ makes the rate proportional to the logarithm ratio of
the two, $\ln(k/g_i T)$, giving rise to unphysical, negative rates when extrapolated to 
sufficiently small momenta. The main ingredient for the DM abundance
is obtained by integrating  the rate over the gravitino momentum 
and thermal distribution. This too turns negative for $g_i\gtrsim 1$,
thus challenging the reliability of this calculation for phenomenology.

To address the issue, \cite{Rychkov:2007uq} introduced a new computational method
which replaces HTL resummation with gauge-dependent one-loop, one-particle-irreducible (1PI)
resummation in the IR-sensitive
parts of the calculation. This method was later extended to axinos in~\cite{Strumia:2010aa} and axions 
in~\cite{Salvio:2013iaa,DEramo:2021psx,DEramo:2021lgb,Becker:2025yvb}; it has been recently
refined in \cite{Eberl:2020fml,Eberl:2024pxr}. It is 
widely thought to be the state of the art for gravitino production calculations.
This method was however shown recently by two of us in \cite{Bouzoud:2024bom} to unavoidably
give rise to pathologically divergent, unphysical rates in non-abelian gauge theories. As the divergence arises from a small 
corner of phase space, it must have been missed or artificially, accidentally
regulated by numerical effects in the implementations in 
\cite{Rychkov:2007uq,Strumia:2010aa,Salvio:2013iaa,DEramo:2021psx,DEramo:2021lgb,Eberl:2020fml,Eberl:2024pxr}.

The analysis in \cite{Bouzoud:2024bom} was limited to the axion case. Here we will show that the issue 
affects gravitino production as well and we will extend the computational methods introduced there 
to this case. To this end, we use the \acrs{} code we are releasing~\cite{autosite} and documenting 
in a companion paper~\cite{autotherm} to perform the calculation with full automation 
from the Lagrangian of the MSSM minimally coupled to supergravity---we shall also 
determine the Gravitational Wave (GW or graviton) production rate, confirming earlier 
results in~\cite{Ringwald:2020ist} and commenting on the SUSY-related relations
between the gravitino and graviton rates as predicted by~\cite{Caron-Huot:2008vbk}. 

As we shall see, 
\acrs{} provides three LO-equivalent implementations of 
HTL resummation: the first one is  the strict leading order 
of \cite{Braaten:1991dd,Bolz:2000fu,Pradler:2006qh,Pradler:2006hh,Pradler:2006tpx}.
The gauge-coupling component of our results agrees perfectly with  \cite{Pradler:2006qh,Pradler:2006hh,Pradler:2006tpx};
the top-Yukawa component is in good agreement with \cite{Rychkov:2007uq}.
The two other implementations---or \emph{schemes}---resum 
subsets of potentially large order-$g_i$ thermal corrections
from soft bosons. In particular, we shall show how the \emph{tuned scheme} 
is constructed so as to yield a positive-definite rate for all $k$ and to 
agree with the strict LO rate for $k\gg g_i T$ at small $g_i$, while at the same 
time avoiding the gauge-dependent pitfalls we mentioned. 
The spread between these different schemes will then be used as a first 
estimate of the theory uncertainty of the calculation of the gravitino rate and abundance.

The paper is organized as follows: in sec.~\ref{sec:thprod} we lay 
down the field-theoretical basis of the calculation, 
to then show in sec.~\ref{sec:auto}
how it is tackled with \acrs{}, showing pedagogical intermediate results 
and discussing the SUSY relation with the graviton rate. 
Section~\ref{sec:res}
is then dedicated to a discussion of our numerical results for the gravitino
rate, which we also package in a compact parameterization extending 
that of~\cite{Ellis:2015jpg}. The code used to obtain all results and plots
is available within the \acrs{} release. 
We draw our conclusions in sec.~\ref{sec:concl}. 
Appendix~\ref{app_slava} is dedicated to an in-depth analysis of the 
gauge-dependent resummation of~\cite{Rychkov:2007uq}, proving how it leads 
to an ill-defined, divergent rate.

%
\section{Definition of the rate}
\label{sec:thprod}

We are interested in the calculation in the $T\gg m_{3/2}$ limit: if the 
universe were to reheat at $T\lesssim m_{3/2}$ thermal production would 
be Boltzmann-suppressed. 
In this high temperature limit, 
we can
use the gravitino/Goldstino equivalence illustrated in detail in~\cite{Rychkov:2007uq}.
In a nutshell, similarly to what happens when taking the zero-mass limit 
for a spin-two particle~\cite{vanDam:1970vg,Zakharov:1970cc}, 
the gravitino would not reduce simply to its 
two $m_s=\pm 3/2$ spin states; a so-called Goldstino component
survives, with $m_s=\pm 1/2$.
Let us call $\psi$ the spin-3/2 component of the gravitino ($\tilde G$)
and $\chi$ the spin-1/2 Goldstino component, where 3/2 and 1/2
indicate the modulus of $m_s$. We then define 
$f^{ }_{\alpha}\equiv (2\pi)^3\mathrm{d}N_\alpha/(\mathrm{d}^3\mathbf{k}\mathrm{d}^3\mathbf{x})$ 
as the spin-averaged phase-space distribution for the $\alpha=\psi,\chi$ states.

These distributions evolve according to the generic form~\cite{Bodeker:2015exa}
\begin{equation}
 {\dot {f}^{ }_{\alpha} }
 \; = \; 
 \Gamma_\alpha(k) \, 
 \bigl[
 \nF^{ }(k) - f^{ }_{\alpha}(t,k) 
 \bigr]
 + \rmO\biggl( \frac{1}{m_\rmi{Pl}^4}\biggr)
 \;, \label{rate_gen}
\end{equation}
where $k \equiv |\vec{k}|$, 
$\nF(k) \equiv 1 / (e^{k/T} + 1)$
is the Fermi-Dirac distribution, and 
$ \dot{g}(t,k(t)) \equiv [\partial^{ }_t - H k \partial^{ }_k]\,g(t,k(t))$ 
with $H$ being the Hubble rate and $g$ any function of time and redshifting $k$.
In the ultrarelativistic (UR) limit we are considering, $k\sim T\gg m_{3/2}$,
we also take $k^0\to k$. $\Gamma_\alpha(k)$ is then the thermal production rate
at first order in the gravitational coupling.
The total gravitino number density, which determines the DM abundance, is given by 
\begin{equation}
    \label{number}
    n_{\tilde G} = 2\int \frac{{\rm d}^3\vec{k}}{(2\pi)^3} \bigl( f^{ }_{\psi}+f^{ }_{\chi} \bigr)\,,
\end{equation}
where the factor of 2 account for the two spin states for each component.
Since production happens in a freeze-in regime, $f_\alpha\ll \nF$ everywhere and the 
corresponding $f_\alpha$-proportional \emph{loss term} in Eq.~\eqref{rate_gen}
can be safely neglected.

The Lagrangian describing the $\psi$ and $\chi$ couplings to 
matter and radiation reads, following the notation of~\cite{Rychkov:2007uq}
\begin{equation}
    \label{gravcoupling}
    \mathcal{L}\supset \frac{\kappa}{4}\bar\psi_\mu S^\mu
    +\frac{\kappa}{\sqrt{24}m_{3/2}}\bar\chi\partial_\mu S^\mu+\text{ h.c.}
\end{equation}
where $\kappa\equiv\sqrt{32\pi}/\mpl$ and $S^\mu$ is the \emph{supercurrent}. 
In what follows we shall consider the thermal bath to be described by the MSSM 
at a temperature above that of any superpartner mass. Our goal is to obtain the rate 
at leading order in $\kappa$ and in the couplings of 
the MSSM---speficifically
the gauge couplings $g_i$ and the top Yukawa $h_t$. To this end, as
illustrated clearly in~\cite{Rychkov:2007uq}, one can use the supercurrent 
without considering any SUSY-breaking soft terms for $\psi$ production.
As that current is conserved, one would find from it a vanishing Goldstino rate.
Dimensionality dictates that the leading contribution at temperatures
larger than the SUSY-breaking scale come from soft terms with a coupling 
of mass dimension one, such as fermion masses. In the MSSM these are indeed 
the gaugino masses $M_i$ and the top Yukawa soft term $A_t$. The equivalence
theorem then shows that $\Gamma_\psi$ is simply the $\Gamma_\chi$ rate obtained from these terms with these rescalings 
\begin{equation}
    \label{equiv-rescaling}
    \Gamma^{ }_{\chi}\Big\vert_\mathrm{top}
    \; = \; 
    \frac{A_t^2}{3m_{3/2}^2} \,
    \Gamma^{ }_{\psi}\Big\vert_\mathrm{top}
    \ , 
    \qquad
    \Gamma^{ }_{\chi}\Big\vert_{\mathrm{gauge},i}
    \; = \; 
    \frac{M_i^2}{3m_{3/2}^2} \,
    \Gamma^{ }_{\psi}\Big\vert_{\mathrm{gauge},i}
    \; .
\end{equation}

Finally, we remark that, as pointed out in \cite{Strumia:2010aa}, 
the Goldstino rate is closely related to the axino one. The latter is 
a superpartner of the axion; earlier computations of its thermal production
rate \cite{Brandenburg:2004du} suffered from the same negativity 
problem we mentioned before. Exploiting the proportionality
of the axino and Goldstino couplings to gluons and gluinos, 
the axino rate can be obtained from the $g_3^2$-proportional 
part of $\Gamma_\chi(k)$ by multiplying it by $3\alpha_s^2m_{3/2}^2/(2\pi^2\kappa^2 M_3^2 f_a^2)$,
with $\alpha_s=g_3^2/(4\pi)$ and $f_a$ the axion scale.

%
\section{Computation using \acrs{}}
\label{sec:auto}

The LO determination of the gravitino rate follows closely that of other 
UR states coupled to a bath via dimension-five operator, such as axions or gravitons.
As discussed at length in \cite{autotherm}, \acrs{} automates
the computational method described in detail in \cite{Bouzoud:2024bom} for axions
and in \cite{Ghiglieri:2020mhm} for gravitons---it is also applicable to
states coupled via operators of dimension other than five. \acrs{} starts 
by generating the Feynman rules with \textsc{FeynRules}~\cite{Alloul:2013bka} 
from a user-provided model file.
These are then used with \textsc{FeynArts/FormCalc}~\cite{Hahn:2000kx,Hahn:2016ebn}
to compute and square matrix elements for all $2\to 2$ processes 
$ab\to cd$, with $a,b,c$ thermal constituents and $d$ the state of interest. 
The resulting $|\mathcal{M}_{ab\to cd}|^2$ are then numerically convoluted with thermal 
distributions over the phase space; would-be infrared divergences from 
soft gauge-boson or fermion exchange are cured by HTL resummation automatically
with three different, LO-equivalent schemes.

In what follows we illustrate the main steps of this procedure and comment on 
intermediate results. 

\subsection{Automated computation of the naive rate}
\label{sub:naive}

As a first step, we have  created a \textsc{FeynRules} model file
for the symmetric-phase MSSM coupled to gravitons and Goldstinos. We have
not included the $\psi$ component; though supported by \textsc{FeynRules}, Rarita--Schwinger
fields are not implemented in \textsc{FeynArts/FormCalc} (or in any other generator, to the
best of our knowledge). We will thus exploit the Goldstino equivalence to
obtain $\Gamma_\psi$ from $\Gamma_\chi$. This model file is available within the \acrs{}
source release; it can also be accessed through the online documentation~\cite{autosite}.

Provided with this model file, \acrs{} generates the Feynman rules
and evaluates all needed matrix element squared and thermal masses, for later use.
On an M3 Pro Apple laptop with Mathematica 14 these operations take approximately 3.5 minutes.
We present here the resulting expression for the rate, before any handling 
of IR divergences. To this end, we call it \emph{naive}, as it arises from 
a direct application of the LO Boltzmann picture. The naive $\psi$ rate then reads
\begin{eqnarray}
 \hspace{-0.8cm} 
 \Gamma_{\psi}(k)\bigg\vert_\text{naive} \!\! &=&  \!\! 
   \frac{\kappa^2}{4k\,\nF^{ }(k)}
   \int \! {\rm d}\Omega^{ }_{2\to2} \bigg\{
   \nonumber\\
   \label{allfermions}
   &+&\nF^{ }(p_1)\,\nF^{ }(p_2)\,[1-\nF^{ }(k_1)] \bigg[
   C^\psi_\mathrm{gauge}
   \left(\frac{s t}{u}+\frac{s u}{t}+\frac{t u}{s}\right)\bigg]\\
   &-&\nB^{ }(p_1)\,\nF^{ }(p_2)\,[1+\nB^{ }(k_1)]
   \bigg[  C^\psi_\mathrm{Yukawa}\;  t 
   +\frac{C^\psi_\mathrm{gauge}}{2}
   \frac{s^2+u^2}{t}\bigg]\label{bfb}\\
 &-&\nF^{ }(p_1)\,\nB^{ }(p_2)\,[1+\nB^{ }(k_1)]
   \bigg[   C^\psi_\mathrm{Yukawa}\; u 
   +\frac{C^\psi_\mathrm{gauge}}{2}
   \frac{s^2+t^2}{u}\bigg]\label{fbb}\\
   &+&\nB^{ }(p_1)\,\nB^{ }(p_2)\,[1-\nF^{ }(k_1)]
   \bigg[  C^\psi_\mathrm{Yukawa}\; s 
   +\frac{C^\psi_\mathrm{gauge}}{2}
   \frac{t^2+u^2}{s}\bigg]\label{bbf}\bigg\}
  \;. \hspace*{4mm} \la{boltzmann}
\end{eqnarray}
where 
$\nB(p) \equiv [\exp(p/T)-1]^{-1}_{ }$ is the Bose--Einstein distribution, 
$\nF(p) \equiv [\exp(p/T)+1]^{-1}_{ }$ is the Fermi--Dirac distribution, 
and  $\int \! {\rm d}\Omega^{ }_{2\to2}$
is the Lorentz-invariant $2\to2$ phase space
for $\mathbf{p}_1,\mathbf{p}_2$ incoming states
and $\mathbf{k}_1,\mathbf{k}$ outgoing ones. 
All momenta except $\vec{k}$ are 
integrated over with the measure 
\begin{equation}
  \label{phase_space}
  \int \! {\rm d}\Omega^{ }_{2\leftrightarrow 2}
  \equiv 
  \int \! \frac{{\rm d}^3\vec{p}_1^{ }
             \,{\rm d}^3\vec{p}_2^{ } 
             \,{\rm d}^3\vec{k}_1^{ }
              }
         {           (2 {p}^{ }_1)
                   \,(2 {p}^{ }_2)
                   \,(2 {k}^{ }_1)
           \, (2\pi)^9  }
  \, (2\pi)^4\, 
  \delta^{(4)}_{ }
  \bigl( \P^{ }_1 + \P^{ }_2 - \K^{ }_1 - \K \bigr),
\end{equation}
where uppercase 
calligraphic
letters denote Minkowski four-vectors, 
 i.e. $\P = (p^0_{ }, \vec{0})$. 
(For notational ease, we will often employ $p^{ }_0$ instead of $p_{ }^0$ for the energy variable.) 
The usual Mandelstam variables are defined by 
\begin{equation}
  \label{eq:stu}
  s 
  \; = \; 
  (\P^{ }_1+\P^{ }_2)^2
  ,\quad
  t
  \; = \;
  (\P^{ }_1-\K^{ }_1)^2
  ,\quad
  u
  \;=\;
  (\P^{ }_2-\K^{ }_1)^2
  ,
\end{equation}
and the $C^\psi_{ }$ coefficients read
\begin{equation}
  \label{mssmcoeffs}
  C^\psi_\mathrm{gauge}
  \; \equiv \;
  11 g_1^2+
  27g_2^2+
  72g_3^2
  \; , \qquad 
  C^\psi_\mathrm{Yukawa}
  \; \equiv \; 
  11g_1^2+
  21g_2^2+
  48g_3^2+9 |h^{ }_t|^2
  \; . 
\end{equation}
They stem from MSSM gauge-boson and Yukawa couplings respectively.\footnote{%
By noting that, for massless external states
\begin{equation}
  \frac{s t}{u}+
  \frac{s u}{t}+
  \frac{t u}{s}
  \; = \;
  -\frac12 \bigg(
    \frac{s^2+t^2}{u}+
    \frac{s^2+u^2}{t}+
    \frac{t^2+u^2}{s}
  \bigg),
\end{equation}
the symmetric structure of the gauge couplings appears more clearly 
and so does the lack 
of a Yukawa contribution in \eq\eqref{allfermions}.
}
As mentioned, the Goldstino rate  $\Gamma_{\chi}(k)$ is obtained 
from  $\Gamma_{\psi}(k)$
with the replacements $C^\psi\to C^\chi$, where
\begin{eqnarray}
  C^\chi_\mathrm{gauge}
  & = & 
  \frac{
    11g_1^2M_1^2+
    27g_2^2M_2^2+
    72g_3^2M_3^2
  }{
    3m_{3/2}^2
  }
  \; , \nonumber\\[2mm] 
  C^\chi_\mathrm{Yukawa}
  & = & 
  \frac{
    11g_1^2M_1^2+
    21g_2^2M_2^2+
    48g_3^2M_3^2+
    9 |h^{ }_t|^2A_t^2
  }{
    3m_{3/2}^2
  } 
  \; .
  \label{mssmcoeffschi}
\end{eqnarray}

%
\begin{figure}[t]
\centerline{
  \includegraphics[width=0.4\linewidth,trim=0 0 0 3.cm, clip]{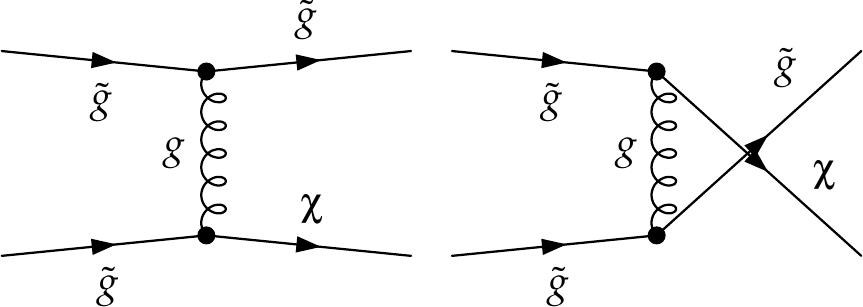}
}
\vspace{-2mm}
\caption[a]{\small
  Tree-level diagrams contributing to the 
  $\tilde g\tilde g\to \tilde g\chi$ process 
  (generated automatically with \acrs{}). 
  Here $\chi$ is the spin-$1/2$ Goldstino, 
  $g$ represents an ordinary gluon, and $\tilde g$ a gluino.
}
\label{fig:gluinos}
\end{figure}
%

The numerical factors can be easily understood~\cite{Pradler:2006tpx,Rychkov:2007uq,Eberl:2020fml,Eberl:2024pxr}: 
11, 21 and 48 are the sum, over all
involved chiral multiplets, of their traced quadratic Casimir operators. E.g. in 
the SU(3) case we have the $Q_L$, $U_R$ and $D_R$ multiplets, giving 
$d_F C_F(6+3+3)=12 T_F d_A=48$, where $d_F C_F$ is the traced quadratic fundamental Casimir,
which occurs $2\times 3$ times for $Q_L$ (SU(2) and generation) and 3 times for the other two multiplets.
The coefficients $27g_2^2$ and $72g_3^2$ appearing for the non-abelian groups in $C^\psi_\mathrm{gauge}$ 
are obtained by adding to this sum the contribution of the 
gauge multiplet, which gives an extra factor of $d_A C_A$, which equals 6 and 24 respectively.
We remark that \acrs{} outputs not just
human- and machine-readable versions of Eqs.~\eqref{allfermions}--\eqref{bbf} but also
expressions for the matrix elements squared for each contributing process and the corresponding Feynman 
diagrams. For instance, 
\begin{equation}
    \label{gluinos}
    \bigl\vert\mathcal{M}_{\tilde g\tilde g\to \tilde g\chi}\bigr\vert^2=
    \frac{4g_3^2\kappa^2 M_3^2}{m_{3/2}^2}\bigg(2s+\frac{su}{t}+\frac{st}{u}\bigg)
    =
    \frac{4g_3^2\kappa^2 M_3^2}{m_{3/2}^2}\bigg(-\frac{s^2}{t}-\frac{s^2}{u}\bigg)\,,
\end{equation}
where $\tilde g$ are the gluinos. This expression is summed over all degeneracies of the external states. In
our \acrs{} implementation of the model, we have exploited the $T\gg m$ limit to 
describe Majorana fermions such as $\tilde g$ and $\chi$ as Weyl ones. 
Within \acrs{} the two helicity states of a Weyl field are treated as distinct particle and 
antiparticle modes, so $\tilde g$ and $\chi$ have one single helicity state in Eq.~\eqref{gluinos}.
The corresponding Feynman diagrams, as created by \acrs{} through \textsc{FeynArts}, are
shown in Fig.~\ref{fig:gluinos}. 

Our expression in Eqs.~\eqref{allfermions}--\eqref{bbf} agrees with those in 
\cite{Bolz:2000fu,Pradler:2006hh,Pradler:2006qh,Pradler:2006tpx,Rychkov:2007uq,Eberl:2020fml,Eberl:2024pxr}, except
for a discrepancy in the $|h^{}_t|^2$-proportional term in \cite{Eberl:2024pxr}. We shall return to
this later. We now proceed to handle the emergence of collective effects, which is the distinguishing feature
of \acrs.

%
\subsection{Automated handling of HTL resummation}
\label{sub:HTL}

The rate in Eqs.~\eqref{allfermions}--\eqref{bbf}, taken as is, diverges logarithmically
in the infrared, due to the $t$ and $u$ denominators. These correspond to gauge-boson mediated
processes such as those shown in Fig.~\ref{fig:gluinos}. This Coulomb-like divergence
signals that the exchanged vector boson fails to resolve
individual hard ($p\sim T$) bath constituents at sufficiently low values of $t$ or $u$. 
It is only by properly accounting for the emergence of collective effects at
the \emph{screening scale} $g_i T$ that a finite rate is recovered.

The LO-correct rate requires HTL resummation for vector boson momenta $q\lesssim g_i T$. The prototype
calculation for abelian axion production in Ref.~\cite{Braaten:1991dd} showed how to split the calculation in two regions 
based on the magnitude of $q$ by introducing 
an intermediate scale $q^*$ satisfying 
$g_i T \ll q^* \ll T$ but being otherwise arbitrary. 
One then uses HTL resummation 
only for $q \leq q^*$, and standard kinetic theory otherwise. 
This yields a \emph{strict LO rate} for $k\gg g_i T$, 
which extrapolates to negative 
values for $k\lesssim g_i T$. 
This method was adopted in the calculations of 
Refs.~\cite{Bolz:2000fu,Pradler:2006hh,Pradler:2006qh,Pradler:2006tpx}.

\acrs{}  automates this calculation, as well as that of two LO-equivalent schemes which 
resum a partial subset of higher-order effects. We now illustrate the main steps
of the calculation, showing how the different schemes emerge.
First, \acrs{} exploits the fact that one can freely relabel 
the initial momenta, $\mathbf{p}^{ }_1\leftrightarrow \mathbf{p}^{ }_2$ and correspondingly
$t\leftrightarrow u$. This reduces Eqs.~\eqref{allfermions}--\eqref{bbf} into
\begin{align}
 \Gamma_{\psi}(k)\bigg\vert_\text{naive}  = 
   \frac{\kappa^2}{4k\,\nF^{ }(k)}
   \int \! {\rm d}\Omega^{ }_{2\to2} \bigg\{
   -&\nF^{ }(p_1)\,\nF^{ }(p_2)\,[1-\nF^{ }(k_1)] \bigg[
   C^\psi_\mathrm{gauge}
   \left(\frac{s^2+u^2}{t}+\frac{t^2}{s}\right)\bigg]\nonumber \\
   -&\nB^{ }(p_1)\,\nF^{ }(p_2)\,[1+\nB^{ }(k_1)]
   \bigg[  2\,C^\psi_\mathrm{Yukawa}\;  t 
   +C^\psi_\mathrm{gauge}
  \frac{s^2+u^2}{t}\bigg]\nonumber \\
   +&\nB^{ }(p_1)\,\nB^{ }(p_2)\,[1-\nF^{ }(k_1)]
   \bigg[  C^\psi_\mathrm{Yukawa}\; s 
   +C^\psi_\mathrm{gauge}
   \frac{t^2}{s}\bigg]\bigg\}\,. \label{reshuffle}
\end{align}
In this form, it appears clearly that only the $(s^2+u^2)/t$-proportional 
parts will be naively IR divergent and in need of resummation. The remainder 
is finite; it reads
\begin{eqnarray}
 \Gamma_{\psi}(k)\bigg\vert_\text{finite}  \!\! &=&  \!\! 
   \frac{\kappa^2}{4k\,\nF^{ }(k)}
   \int \! {\rm d}\Omega^{ }_{2\to2} \bigg\{
   C^\psi_\mathrm{gauge}
  \frac{t^2}{s} \bigg[\nB^{ }(p_1)\,\nB^{ }(p_2)-\nF^{ }(p_1)\,\nF^{ }(p_2)
   \bigg]\,[1-\nF^{ }(k_1)]\nonumber \\
  &&+\; C^\psi_\mathrm{Yukawa}\,\nB^{ }(p_1)\,
   \bigg[   s\;\nB^{ }(p_2)\,[1-\nF^{ }(k_1)] 
   -2t\;\nF^{ }(p_2)\,[1+\nB^{ }(k_1)]\bigg]\bigg\}
  \;. \hspace*{4mm} \label{finite}
\end{eqnarray}
We see how the $C^\psi_\mathrm{gauge}$ part in this expression would vanish
in the limit of Maxwell--Boltzmann statistics. We shall
explain later how \acrs{} deals with the numerical integration of Eq.~\eqref{finite}.

Let us now tackle the would-be divergent part. Two of the schemes implemented by \acrs{},
namely the strict LO and the \emph{subtracted} one rely on, as the latter name suggests,
a subtraction of the IR-divergent, soft-gluon limit. This limit is then 
added back with HTL resummation and evaluated analytically with the
analyticity-based methods of \cite{Aurenche:2002pd,CaronHuot:2008ni}. 
In the subtracted case this 
gives---see~\cite{Besak:2012qm,Ghiglieri:2016xye,Ghiglieri:2020mhm,Bouzoud:2024bom,autotherm}
\begin{eqnarray}
 \Gamma_{\psi}(k)\bigg\vert^t_\text{subtr} \!\! &=&\!\! -
   \frac{\kappa^2 C^\psi_\mathrm{gauge}}{4(4\pi)^3_{ }k^2}
 \int^k_{-\infty} \! {\rm d}q^{ }_0 
 \int^{2k - q^{ }_0}_{|q^{}_0|} \! {\rm d}q 
 \int_{-\pi}^{\pi} \! \frac{{\rm d}\varphi}{2\pi} \bigg\{
   \nonumber\\
   &+&
  \int_{q_+^{ }}^\infty \! {\rm d}p^{ }_1  \frac{s^2+u^2}{t}\frac{\nF^{ }(k{-}q_0)}{\nF^{ }(k)}\,\bigg[\nF^{ }(p_1)\,[1-\nF^{ }(p_1{-}q_0)]
   +\nB^{ }(p_1)\,[1+\nB^{ }(p_1{-}q_0)]\bigg]\nonumber\\
   &-&\int_{0}^\infty \! {\rm d}p^{ }_1 \frac{8k^2p^{ 2}_1t
   }{q^4_{ }}(1 
  -\cos\varphi\,)^2\,\bigg[\nF^{ }(p_1)\,[1-\nF^{ }(p_1)]
   +\nB^{ }(p_1)\,[1+\nB^{ }(p_1)]\bigg]
 \bigg\}\nonumber\\
 &+&\frac{ \kappa^{2}T}{32 \pi} \sum_{i=1}^3 d_i \mDi^2 \ln\left(1+\frac{4 k^{2}}{\mDi^2}\right)\,.\label{subtrlog}
\end{eqnarray}
Here $q_0$ and $q$ are the energy and momentum of the exchanged gluon, $q_\pm\equiv(q_0\pm q)/2$ 
and $\varphi$ is the azimuthal angle between  $\mathbf{p}^{ }_1$ and $\mathbf{k}^{ }$. 
The Mandelstam invariants are given in terms of the 
integration variables as
\be
 t 
   = q^2_0 - q^2_{ }
 \;, \quad
 u
   = 2\,(\mathbf{k}\cdot\mathbf{p}^{ }_1 
                   - k p^{ }_{1})
 \;, \quad
  s 
   = - t - 2\,(\mathbf{k}\cdot\mathbf{p}^{ }_1 
                   - k p^{ }_{1})
 \;. \la{mandelstam_2to2}
\ee
From the definition of $\varphi\,$, we may write
\ba
 \mathbf{k} \cdot \mathbf{p}^{ }_1 = 
 k p^{ }_1 
 \, 
 \big[
 \cos\theta^{ }_{\mathbf{q},\mathbf{k}} 
 \cos\theta^{ }_{\mathbf{q},\mathbf{p}_1}
 + 
 \, 
 \sin\theta^{ }_{\mathbf{q},\mathbf{k}}
 \sin\theta^{ }_{\mathbf{q},\mathbf{p}_1}
 \, \cos\varphi
 \, \big]
 \;, 
\ea
where the remaining two angles 
are fixed as 
$\cos\theta^{ }_{\mathbf{q},\mathbf{k}} 
= \frac{q_0^{ }}{q} - \frac{t}{2 q k}$ 
and
$\cos\theta^{ }_{\mathbf{q},\mathbf{p}_1} 
= \frac{q_0^{ }}{q} - \frac{t}{2 q p^{ }_1}$.

The third line in \eq\eqref{subtrlog} is the subtraction term, obtained by expanding the second line
for $k,p_1\gg q_0,q$ at first order. The final line is precisely the 
HTL-resummed evaluation of the subtracted part. 
Here $d_i=(1,3,8)$ are the
multiplicities associated with each gauge group and $\mDi$ the corresponding Debye masses.
\acrs{} determines these automatically, finding in the case of the MSSM
\begin{equation}
    \label{debyemasses}
    m_{\rmii{D}1}^2 \; = \; \frac{11}{2}g_1^2 T^2\,,\qquad 
    m_{\rmii{D}2}^2 \; = \; \frac{9}{2}g_2^2 T^2\,,\qquad 
    m_{\rmii{D}3}^2 \; = \; \frac{9}{2}g_3^2 T^2\,,
\end{equation}
in agreement with the known values in Ref.~\cite{Comelli:1996vm}. We remark that 
the third line of \eq\eqref{subtrlog} shows how SUSY causes an equal number
of fermionic and bosonic degrees of freedom to contribute to the Debye masses.
These are the $\nF(1-\nF)$ and $\nB(1+\nB)$ terms respectively. Quantum statistics, though,
causes their contribution to differ, since 
$\int_0^\infty dp_1\,p_1^2\,\nB(p_1)[1+\nB(p_1)]=2T\int_0^\infty dp_1\,p_1\,\nB(p_1)=\pi^2T^3/3$
is twice its fermionic counterpart. 
This explains why the coefficients multiplying
$g_i^2 T^2$ in \eq\eqref{debyemasses} are $1/(2d_i)$ 
times the corresponding ones in $C^\psi_\mathrm{gauge}$.

The strict LO is obtained similarly by 
the replacement~\cite{Bouzoud:2024bom,autotherm}
\begin{equation}
    \label{strictlog} \Gamma_{\psi}(k)\bigg\vert^t_\text{strict LO}= \Gamma_{\psi}(k)\bigg\vert^t_\text{subtr}
    -\frac{ \kappa^{2}T}{32 \pi} \sum_{i=1}^3 d_i \mDi^2 \ln\left(1+\frac{4 k^{2}}{\mDi^2}\right)
    +\frac{ \kappa^{2}T}{32 \pi} \sum_{i=1}^3 d_i \mDi^2 \ln\left(\frac{4 k^{2}}{\mDi^2}\right).
\end{equation}
This shows clearly how the subtracted scheme has resummed, through the unitary constant in the logarithm,
a subset of higher-order terms in $\mDi^2/(4k^2)$. 
This also highlights why these calculations become extrapolations once $k\lesssim \mDi$. 

The third scheme in \acrs{} is constructed, following Ref.~\cite{Peshier:2008bg,Gossiaux:2008jv,York:2014wja},
by introducing in the would-be
$1/q^4$ denominators---see the third line in \eq\eqref{subtrlog}---a
constant mass $\xi \mDi$, i.e.
\begin{eqnarray}
 \Gamma_{\psi}(k)\bigg\vert^t_\text{tuned} \!\! &=&\!\! -\sum_{i=1}^3  
   \frac{\kappa^2C_{\mathrm{gauge}\,i} }{4(4\pi)^3_{ }k^2}
 \int^k_{-\infty} \! {\rm d}q^{ }_0 
 \int^{2k - q^{ }_0}_{|q^{}_0|} \! {\rm d}q 
 \int_{q_+^{ }}^\infty \!   {\rm d}p^{ }_1 
 \int_{-\pi}^{\pi} \! \frac{{\rm d}\varphi}{2\pi} \frac{\nF^{ }(k-q^0)}{\nF^{ }(k)}
   \nonumber\\
   &&\times
 \,\bigg[\nF^{ }(p_1)\,[1-\nF^{ }(p_1-q^0)]
   +\nB^{ }(p_1)\,[1+\nB^{ }(p_1-q^0)]\bigg]\nonumber\\
   &&\times\bigg[\frac{q^4}{2(q^2+\xi^2\mDi^2 )^2}\frac{(s-u)^2}{t}+\frac{t}{2}\bigg]\,.\label{tuned}
\end{eqnarray}
$\xi=e^{1/3}/2$ has been determined analytically in 
Refs.~\cite{Bouzoud:2024bom,Boguslavski:2023waw}
so that, at small $\mDi/k$, 
$ \Gamma_{\psi}(k)\big\vert^t_\text{tuned}- \Gamma_{\psi}(k)\big\vert^t_\text{strict}=\mathcal{O}(\mDi^3)$.
For this reason, this scheme is called \emph{tuned}. 

For all three schemes, the final result is given by adding the finite part, \eq\eqref{finite}, to 
\eqs\eqref{subtrlog}, \eqref{strictlog} and \eqref{tuned} respectively. 
In all cases \acrs{}
automatically separates finite and would-be divergent parts. The $\varphi$ and $p_1$
integrations, as well as their analogues in \eq\eqref{finite}, are carried out analytically. The
remaining two-dimensional integration in $q_0$ and $q$ is carried out numerically. The mapping 
from \eqs\eqref{allfermions}--\eqref{bbf} to \eqs\eqref{finite}, \eqref{subtrlog}, \eqref{strictlog} and \eqref{tuned}
takes less than 2s on an M3 Pro Apple laptop. Evaluating numerically the 2D integrals for all
three schemes for 100 values of $k/T$ takes the same amount of time on that machine. We will present
the results in sec.~\ref{sec:res}.

\subsection{Relation to the GW rate}
\label{sec_gw_susy} 
We have used the \acrs-based methodology just outlined to compute the 
graviton rate in the MSSM---see~\cite{Ghiglieri:2020mhm} for 
the definition. The naive rate reads, in  agreement with~\cite{Ringwald:2020ist}
\begin{eqnarray}
 \hspace{-0.8cm} 
 \Gamma_G(k)\bigg\vert_\text{naive} \!\! &=& \!\! 
   \frac{\kappa^2}{8k\,\nB^{ }(k)}
   \int \! {\rm d}\Omega^{ }_{2\to2} \bigg\{
   \nonumber\\
   \label{allbosons}
   &+&\nB^{ }(p_1)\,\nB^{ }(p_2)\,[1+\nB^{ }(k_1)] 
  \bigg[ 2 C^\psi_\mathrm{gauge}
   \left(\frac{s t}{u}+\frac{s u}{t}+\frac{t u}{s}\right)\bigg]\\
   &-&\nF^{ }(p_1)\,\nB^{ }(p_2)\,[1-\nF^{ }(k_1)]
   \bigg[ 2 C^\psi_\mathrm{Yukawa}\; t 
   +C^\psi_\mathrm{gauge}\,
   \frac{s^2+u^2}{t}\bigg]\label{fbf}\\
 &-&\nB^{ }(p_1)\,\nF^{ }(p_2)\,[1-\nF^{ }(k_1)]
   \bigg[ 2 C^\psi_\mathrm{Yukawa}\; u 
   +C^\psi_\mathrm{gauge}\,
   \frac{s^2+t^2}{u}\bigg]\label{bff}\\
   &+&\nF^{ }(p_1)\,\nF^{ }(p_2)\,[1+\nB^{ }(k_1)]
   \bigg[ 2 C^\psi_\mathrm{Yukawa}\; s 
   +C^\psi_\mathrm{gauge}\,
   \frac{t^2+u^2}{s}\bigg]\label{ffb}\bigg\}
  \;. \hspace*{4mm} \la{boltzmanngrav}
\end{eqnarray}
where the extra factor of 2 in the denominator of the prefactor
arises from the polarisation degeneracy of gravitons. We
can however observe that it is compensated by polarisation-summed 
matrix elements squared which are twice as large
as the corresponding ones for $\vert m_s\vert=3/2$  gravitinos. Indeed, as expected from
supersymmetry~\cite{Caron-Huot:2008vbk}, the naive graviton and $\psi$-component gravitino rates  
transform into each other by exchanging the statistics of all 4 external states. Hence, 
the high-energy, i.e. $k\gg T$, limits of the two rates agree: 
there one can approximate the 
initial-state distributions by their Maxwell--Boltzmann form and neglect final-state Pauli blocking 
or Bose enhancement, leading to identical expressions for the two members of the same
multiplet. This is borne out also after HTL resummation.

For illustration, let us reproduce here the SM equivalent 
of \eq\eqref{boltzmanngrav} from Ref.~\cite{Ghiglieri:2020mhm}. 
It reads
\begin{eqnarray}
 \hspace{-5mm}
 \Gamma_G(k)\bigg\vert^\mathrm{SM}_\text{naive} \!\! &=& \!\! 
   \frac{\kappa^2}{8k\,\nB^{ }(k)}
   \int \! {\rm d}\Omega^{ }_{2\to2} \bigg\{
   \nonumber\\
   \label{allbosonsSM}
   &+&\nB^{ }(p_1)\,\nB^{ }(p_2)\,[1+\nB^{ }(k_1)] 
    \bigl(g_1^2+15g_2^2+48g_3^2\bigr)
   \left(\frac{s t}{u}+\frac{s u}{t}+\frac{t u}{s}\right)\\
   &-&\nF^{ }(p_1)\,\nB^{ }(p_2)\,[1-\nF^{ }(k_1)]
   \bigg[ 6 |h^{ }_t|^2 t 
   +\bigl(10g_1^2+18g_2^2+48g_3^2\bigr)
   \frac{s^2+u^2}{t}\bigg]\label{fbfSM}\\
   &-&\nB^{ }(p_1)\,\nF^{ }(p_2)\,[1-\nF^{ }(k_1)]
   \bigg[ 6 |h^{ }_t|^2 u 
   +\bigl(10g_1^2+18g_2^2+48g_3^2\bigr)
   \frac{s^2+t^2}{u}\bigg]\label{bffSM}\\
   &+&\nF^{ }(p_1)\,\nF^{ }(p_2)\,[1+\nB^{ }(k_1)]
  \bigg[ 6 |h^{ }_t|^2 s 
   +\bigl(10g_1^2+18g_2^2+48g_3^2\bigr)
   \frac{t^2+u^2}{s}\bigg]\label{ffbSM}\bigg\}
  \,. \hspace*{4mm} \la{boltzmannSM}
\end{eqnarray}
The MSSM completion of the SM result can be understood as arising from more degrees of freedom and
more couplings. The former are responsible for, e.g., changing the coefficients of the four-boson
terms in \eqs\eqref{allbosons} and \eqref{allbosonsSM}. In the SM the SU(3) component 
is $2d_A C_A=48$ from $gg\to gG$ processes. In the MSSM this is complemented by the squarks. Through 
processes like $\tilde q g\to \tilde q G$ and its crossings, they contribute $2 d_F C_F( 6+3+3)=96$, 
where the $6+3+3$ accounts for $\tilde q_L$, $\tilde u_R$ and
$\tilde d_R$ respectively.
The new couplings are the Yukawa ones between gauginos, quarks (leptons) and squarks (sleptons), which 
give rise to terms with the same structure as that originating from the top Yukawa in the SM.

%
\section{Numerical results}
\label{sec:res}

%
\begin{figure}[t]
\centerline{
  \includegraphics[width=0.8\linewidth]{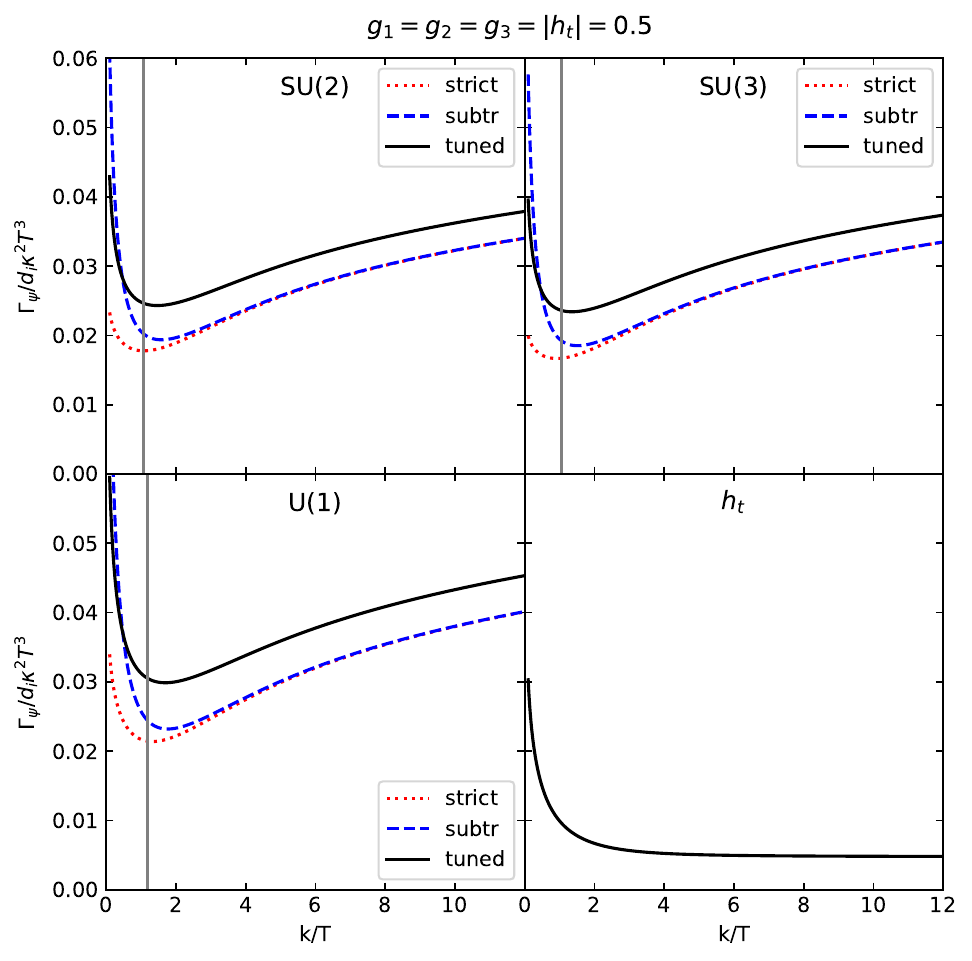}
}
\vspace{-3mm}
\caption[a]{\small
  The three gauge and the top Yukawa contributions 
  $\Gamma_\psi(k)/(d_i\kappa^2T^3)\,$. 
  The four couplings have been artificially set to 
  $g_1=g_2=g_3=|h_t|=0.5$. 
  The gray vertical line indicates where $k=\mDi\,$.
}
\label{fig:smallg}
\end{figure}
%

\begin{figure}[t]
\centerline{
  \includegraphics[width=0.8\linewidth]{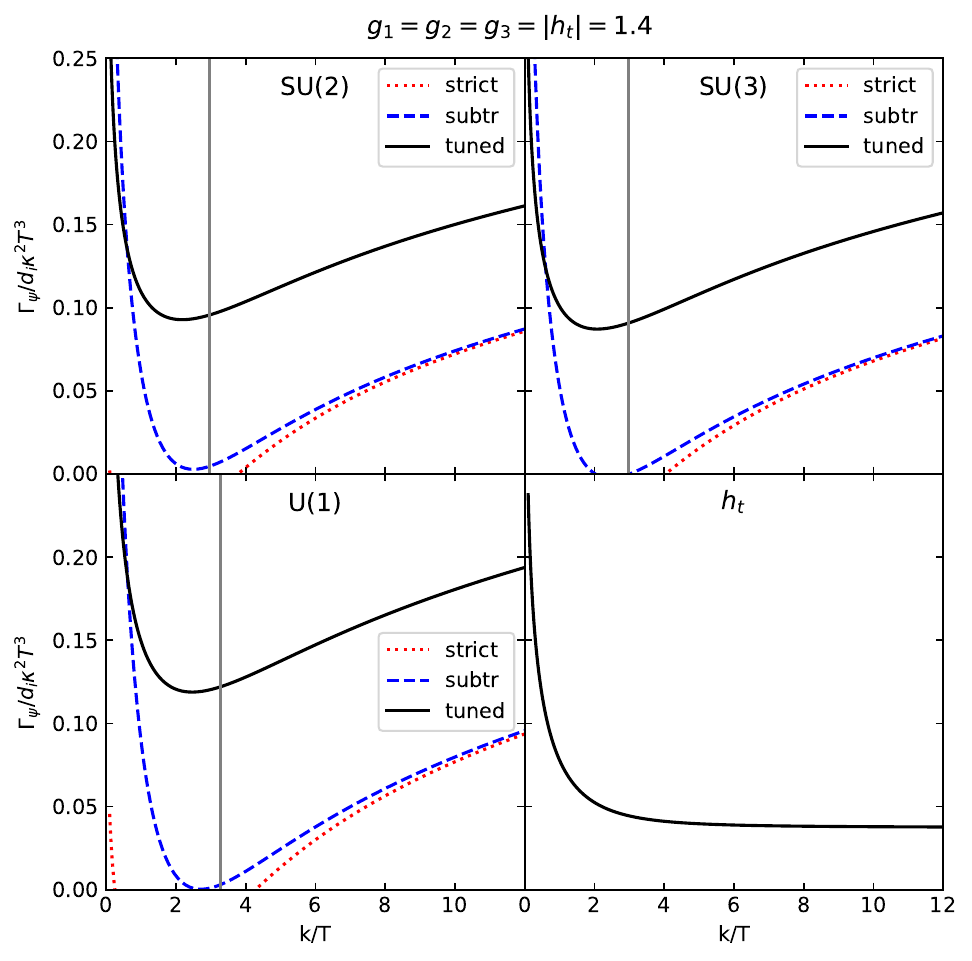}
}
\vspace{-3mm}
\caption[a]{\small
  The rate $\Gamma_\psi(k)/(d_i\kappa^2T^3)\,$, 
  as in \fig\ref{fig:smallg} for the couplings 
  $g_1=g_2=g_3=|h_t|=1.4\,$.
}
\label{fig:largeg}
\end{figure}

Let us start by analyzing the results for $\Gamma_\psi(k)$ as a function of $k/T$
and of the couplings, without for the moment fixing the latter as a function of the temperature.
We do so in \figs\ref{fig:smallg} and \ref{fig:largeg}, where we plot
the contributions to the gravitino rate from the three gauge interactions and the top Yukawa,
$\Gamma_\psi(k)/(d_i\kappa^2T^3)$,  with $d_i=(1,3,8)$ for the three gauge couplings and $d_t=1$. 
The two figures differ only in the choice of the couplings, artificially made equal with 
$g_1=g_2=g_3=|h_t|=0.5$ in \fig\ref{fig:smallg} and $g_1=g_2=g_3=|h_t|=1.4$ in \fig\ref{fig:largeg}.
Note that the Goldstino rate is easily obtained through 
$\Gamma_\psi(k)/(d_i\kappa^2T^3)=\Gamma_\chi(k)\,3m_{3/2}^2/(d_i\kappa^2M_i^2T^3)$.

The small couplings of \fig\ref{fig:smallg} show that for $k>\mDi$, i.e. right of
the gray vertical lines, the strict LO and subtracted rates are very close to each other,
as they differ by order $\mDi^2/k^2\ll 1$ there. The tuned rate there is about 
10\% larger than the other two. As shown in Ref.~\cite{Bouzoud:2024bom}, it differs
from them by a quantity of relative order-$\mDi/T$. Upon extrapolating into the 
$k\lesssim \mDi$ regime we see how the differences between the three rates become
much larger, as expected.

%
%

At the larger couplings in \fig\ref{fig:largeg} these trends get amplified.
The strict LO and subtracted rates, while still in very good agreement for $k>\mDi$,
completely break down in the IR, where the strict rate becomes unphysical. At
$k\sim 8 T$ the tuned rate is about twice as large as the other two, showing the 
emergence of a significant \emph{theory uncertainty}, to which we will return later.
In the IR the tuned rate remains positive-definite by construction.

We also remark that in \figs\ref{fig:smallg} and \ref{fig:largeg}
the $d_i$-normalised non-abelian contributions are quantitatively very close 
at equal couplings. This is easily understood, as $C^\psi_\mathrm{gauge}/d_i$ is 
identical for SU(2) and SU(3)---see \eq\eqref{mssmcoeffs}. It is only the
numerically smaller contribution from $C^\psi_\mathrm{Yukawa}$ that differs in the two cases.

Let us also note that our chosen values for the coupling might appear quite small
to hot-QCD practitioners, as they correspond to $\alpha_s\approx(0.02,0.16)$ respectively.
In QCD, however, one has $m_{\rmii{D}3}/T=\sqrt{3/2} g_3$ with three light flavors ($T<m_c$) and
$m_{\rmii{D}3}/T=\sqrt{2} g_3$ above the electroweak crossover. In the MSSM 
$m_{\rmii{D}3}/T=\sqrt{9/2} g_3$, substantially pushing the Debye scale into the UV.
Hence we have $m_{\rmii{D}3}/T\approx 1.1$ for the smaller coupling and 
$m_{\rmii{D}3}/T\approx 3.0$ at the larger one. The $\mathcal{O}(1)$ spread between
the tuned and strict results in this latter case is then not surprising.

Let us now integrate the rates in momentum, so as to obtain
the quantity determining the total gravitino number density. 
Taking the time derivative of \eq\eqref{number}, 
and using \eqref{rate_gen} with $f^{ }_\alpha \ll \nF\,$, 
we define the number density rate
\begin{equation}
  \label{gammag}
  \gamma^{ }_{\tilde{G}}
  \; \equiv \;
  \frac{ \partial n^{ }_{\tilde{G}} }{\partial t }
  + 3 H n^{ }_{\tilde{G}}
  \; = \;
  2\int \frac{{\rm d}^3\vec{k}}{(2\pi)^3} \,
  \bigl(\Gamma_{\psi}(k)+\Gamma_{\chi}(k) \bigr)\,
  \nF(k)
  \;.
\end{equation}
For $\Gamma^{ }_\chi$ above, one can use \eq\eqref{equiv-rescaling} 
so that everything is expressed in terms of $\Gamma^{ }_\psi\,$.

The top Yukawa contribution does not require resummations 
and thus depends only multiplicatively on $\vert h_t\vert^2$. 
A straightforward numerical
integration of the \acrs-provided rate then yields
\begin{equation}
  \label{topcontrib}
  \gamma_{\tilde{G}}\bigg\vert_\mathrm{top}
  \; = \; 
  1.2860\;
  \frac{ 9 |h^{ }_t|_{ }^2 \kappa_{ }^2 T_{ }^6}{8\pi^5}
  \left(1+\frac{A_t^2}{3m_{3/2}^2}\right)
  \; .
\end{equation}
Following Ref.~\cite{Rychkov:2007uq}, 
we have normalized this expression by the result
that can be obtained analytically with Maxwell--Boltzmann statistics. Our prefactor of $1.2860$ is in good agreement
with their coefficient of $1.30\,$. Conversely, the result in Ref.~\cite{Eberl:2024pxr} is claimed 
to be 12 times smaller at the level of the matrix elements squared. We thus disagree with this claim and we note that, 
if it were true, the expected SUSY relation with the graviton rate
computed in Ref.~\cite{Ringwald:2020ist} and reproduced by us in \eq\eqref{boltzmanngrav} would be broken. 
Upon multiplying the result of the numerical integration in Ref.~\cite{Eberl:2024pxr} by $12$ we find
a coefficient of $1.2709\,$.\footnote{\label{foot_yukawa}%
  While we cannot conclusively explain the small 
  discrepancy between our numerical result 
  and those of Ref.~\cite{Rychkov:2007uq} and 
  $12$ times those of \cite{Eberl:2024pxr}, 
  we note that the form of the Yukawa term in \eq\eqref{finite} 
  does not allow for further simplifications. 
  Namely, given the different statistics for the $s$- and $t$-proportional 
  pieces, the integral of the ``$-2t$ part'' \emph{is not} 
  equal to that of the ``$s$ part''; 
  this would only happen if the statistical functions multiplying 
  $-2t$ were the same as those multiplying $s$. 
  The latter follows from the identity
  $
  \int \! {\rm d}\Omega^{ }_{2\leftrightarrow 2}(2t+s)(...)=0\,
  $, 
  valid only in terms where the incoming states have 
  the same quantum statistics. 
  It is suggestive to note that if we integrate twice the 
  ``$s$ piece'' we find a numerical coefficient of $1.2709\,$, 
  in agreement with $12$ times the result of Ref.~\cite{Eberl:2024pxr}. 
  If we instead integrate twice the $-2t$ piece we find $1.3010\,$, 
  in agreement with Ref.~\cite{Rychkov:2007uq}. 
}

%
\begin{figure}[t]
\centerline{
  \includegraphics[width=0.9\linewidth]{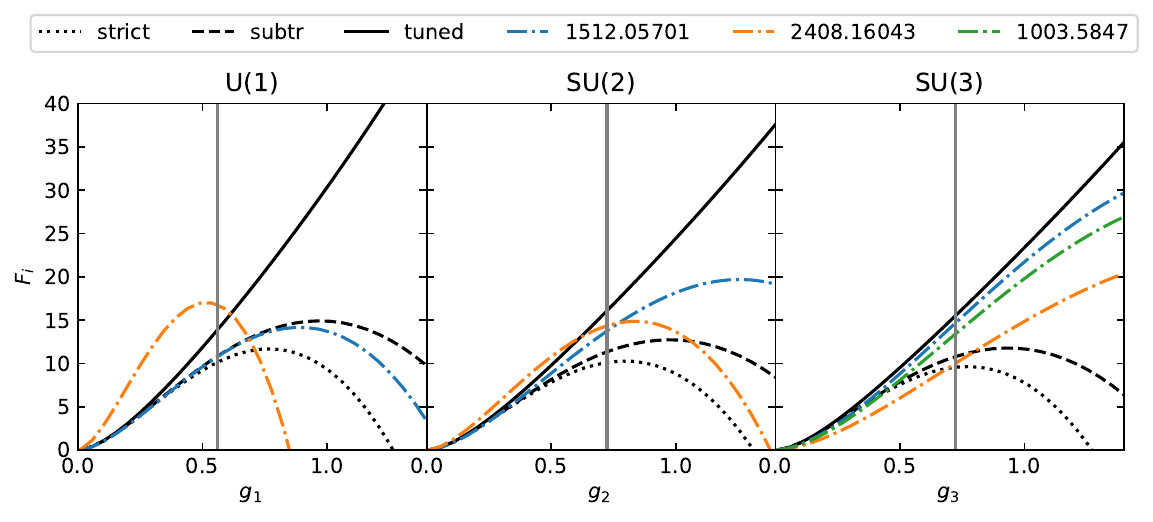}
}
\vspace{-4mm}
\caption[a]{\small
  The rate functions $F_i(g_i)$ for the three gauge groups in
  our three schemes, compared other results 
  in the literature. 
  Specifically, 
    the blue curves are from \eqs(5)--(6) and table~1 of Ref.~\cite{Ellis:2015jpg}; 
    the orange curves are from \eq(4.2) and table~3 of Ref.~\cite{Eberl:2024pxr}; 
    the green curve (rightmost panel only) is \eq(9) of Ref.~\cite{Strumia:2010aa}. 
  The vertical gray lines represent the value of each coupling 
  at the GUT scale,
  namely $5 g_1^2/(12\pi)=g_2^2/(4\pi)=g_3^2/(4\pi)=1/24\,$.
}
\label{fig:fi}
\end{figure}
%

For what concerns the gauge couplings, let us adopt the 
parametrisation of Ref.~\cite{Rychkov:2007uq}, i.e.
\begin{equation}
  \label{gaugecontrib}
  \gamma_{\tilde{G}}\bigg\vert_\mathrm{gauge}
  \; = \; 
  \sum_{i=1}^3\frac{ \kappa_{ }^2 T_{ }^6 }{(4\pi)^3}
  \left(1+\frac{M_i^2}{3m_{3/2}^2}\right)d_i\,F_i(g_i)\,.
\end{equation}
In the case of the strict LO rate, the function can be expressed 
as Refs.~\cite{Pradler:2006hh,Pradler:2006qh,Pradler:2006tpx}
\begin{equation}
  \label{strictLOint}
  F_1^\text{strict}
  \; = \; 
  \frac{ 6\zeta(3) \,m_{\rmii{D}1}^2 }{T^2}
  \ln\frac{1.266}{g_1}
  \, , \ 
  F_2^\text{strict}
  \; = \; 
  \frac{6\zeta(3) \, m_{\rmii{D}2}^2}{T^2}
  \ln\frac{1.312}{g_2}
  \, , \ 
  F_3^\text{strict}
  \; = \;
  \frac{6\zeta(3) \, m_{\rmii{D}3}^2}{T^2}
  \ln\frac{1.271}{g_3},
\end{equation}
where the coefficients of the logarithmic terms in the coupling are
determined analytically. As can be seen from \eq\eqref{strictlog},
the logarithmic dependence on the coupling enters through 
the 
$\frac{\kappa^2 T}{32\pi} d_i^{ } \mDi^2 \ln (1/g_i^2)$ 
piece in $\Gamma_\psi^{ }$. Inserting this
into \eq\eqref{gammag} one the finds
\begin{align}
  \gamma_{\tilde{G}}\bigg\vert_\mathrm{gauge}^\text{LL}
  \; = & \;
  \frac{ \kappa^{2}T }{32 \pi} 
  \sum_{i=1}^3 
  d_i \mDi^2 
  \left(1+\frac{M_i^2}{3m_{3/2}^2}\right)
  2 \int \! \frac{{\rm d}^3\vec{k}}{(2\pi)^3}
  \ln\left(\frac{4 k^{2}}{\mDi^2}\right)\nF(k) 
  \nonumber\\[2mm]
  \; = & \; 
  \frac{ \kappa^{2}T^4}{(4\pi)^3} 
  \sum_{i=1}^3 d_i \mDi^2 
  \left(1+\frac{M_i^2}{3m_{3/2}^2}\right)
  3\zeta(3)
  \Biggl\{
  \ln\left(\frac{4 T^2}{\mDi^2}\right)
  +3-2 \gammaE+\frac{\ln (4)}{3}
  +\frac{2 \zeta '(3)}{\zeta(3)}
  \Biggr\}
  \; , 
  \label{strictlogF}
\end{align}
where LL stands for \emph{leading log}, 
meaning that the $\ln(4T^2/\mDi^2)$ piece within this term captures 
the dominant behavior in the strict asymptotic limit for 
$g^{ }_i \to 0$ such that 
$\ln(1/g^{ }_i)\gg 1$~(and other quantities being fixed).
The constants under the logarithms in \eq\eqref{strictLOint}
arise from the remainder of \eq\eqref{strictlogF} and 
from integrating in \eq\eqref{gammag}
the remainder of the strict LO rate. 
Their numerical values are taken 
from Refs.~\cite{Pradler:2006hh,Pradler:2006qh,Pradler:2006tpx}.

Our numerical results for $F_i^\text{strict}$ 
in \fig\ref{fig:fi} agree perfectly with \eq\eqref{strictLOint}. 
One furthermore sees the expected hierarchy
of the strict LO, subtracted and tuned rate. We further
show through vertical gray lines the value of the coupling at 
the GUT scale, $M_\mathrm{GUT}=2\times 10^{16}$~GeV,    
$5/3 g_1^2/(4\pi)=g_2^2/(4\pi)=g_3^2/(4\pi)=1/24$. 
As the SU(3) (SU(2), U(1)) coupling runs to larger (smaller) 
values with decreasing energy or temperature, 
the relevant regions for phenomenology are to the right
of the gray line for SU(3) and to the left otherwise. 

For illustration, we also compare with Refs.~\cite{Ellis:2015jpg,Strumia:2010aa}, which 
parametrize the results of Ref.~\cite{Rychkov:2007uq}, 
and with the parametrization 
in Ref.~\cite{Eberl:2024pxr} of their own results. These parametrisations
are intended to be valid on the phenomenologically relevant side 
of the gray lines.
We stress that these numerical
results are not to be trusted, as they must arise from artificial, accidental numerical
regularization of a divergent integral, as shown at length in Ref.~\cite{Bouzoud:2024bom}
for the axion rate and as discussed in appendix~\ref{app_slava}. 
Nevertheless,
the relative difference with our most reliable rate, the tuned one, 
is small to moderate
in the validity region of these parameterizations.

%
\begin{table}[t]
\small{
\begin{center}
\setlength{\tabcolsep}{10pt}
\begin{tabular}{c|ccc} 
 $F_i^\text{fit}$  & 
 $c_i$ & 
 $k_i$ & 
 $\delta_i$ 
 \\[1mm]
 \hline
 \\[-4mm] 
  U(1) strict  & 
  11$^\dagger$ & 
  1.266        & 
  0$^\dagger$  
  \\[.5mm]
  U(1) subtr   & 
  11.554       & 
  1.118        & 
  2.841
  \\[.5mm]
  U(1) tuned   & 
  9.987        & 
  1.504        & 
  4.334
  \\[.5mm]
  \hline 
 \\[-4mm] 
  SU(2) strict &
  9$^\dagger$  &
  1.312        &
  0$^\dagger$
  \\[.5mm]
  SU(2) subtr  &
  9.455        & 
  1.163        &
  2.104
  \\[.5mm]
  SU(2) tuned  &
  8.345        & 
  1.497        &
  3.405
  \\[.5mm]
  \hline 
 \\[-4mm] 
  SU(3) strict &
  9$^\dagger$  &
  1.271        &
  0$^\dagger$
  \\[.5mm]
  SU(3) subtr  &
  9.455        & 
  1.128        &
  2.104
  \\[.5mm]
  SU(3) tuned  &
  8.345        & 
  1.446        & 
  3.405
  \\[.5mm]
 \hline 
\end{tabular} 

\vspace{2mm}

{\footnotesize $^{\dagger}$ Value fixed in the fit.}
\end{center}
}
\caption[a]{\small
   Fitted values for the parameters in \eq\eqref{fitdef}. 
   The results for the strict scheme agree with those in Refs.~\cite{Pradler:2006hh,Pradler:2006qh,Pradler:2006tpx}; 
    the $c_i$ and $\delta_i$ coefficients
    are set according to \eq\nr{strictlogF} (dropping the 
    constant next to the logarithm). 
}
\label{tab:fitsfinal}
\end{table}

Our numerical results for $\Gamma_\psi(k)$ are 
easily available through \acrs.
From the provided example files  the numerical rates can be obtained 
rapidly and precisely and integrated in $k$ if needed. Feynman-rule derivation
and the generation and calculation of the matrix elements squared can be
skipped for the problem at hand, as the result is also provided in the example
files for the MSSM. Nevertheless, we also provide a simple 
parametrisation of the $F_i$ functions for readers who are legitimately interested only in it. 
It reads, generalising \eq\eqref{strictLOint} and the form used in 
Refs.~\cite{Pradler:2006hh,Pradler:2006qh,Pradler:2006tpx,Ellis:2015jpg,Eberl:2024pxr}
\begin{equation}
  \label{fitdef}
  F_i^\text{fit}(0.01<g_i<1.4)
  \; = \;
  3\zeta(3)\,g_i^2
  \bigg[c_i\ln\frac{k_i}{g_i}+ \delta_i g_i\bigg]
  \; .
\end{equation}
We note that our definition of $c_i$ differs from that in 
Refs.~\cite{Pradler:2006hh,Pradler:2006qh,Pradler:2006tpx,Ellis:2015jpg,Eberl:2024pxr}, as
the multiplicity $d_i$ is in our case factored 
out of the definition of $F_i$ in \eq\eqref{gaugecontrib}. We have
furthermore added an extra term of relative order $g_i$, 
parametrized by the $\delta_i$ coefficient.
$c_i$, $k_i$ and $\delta_i$ are then fitted to the curves in \fig\ref{fig:fi}; the resulting
values are provided in table~\ref{tab:fitsfinal}. 
Within the \emph{validity region} of the fits $0.01<g_i<1.4$, 
the accuracy is in general at sub-percent level, 
worsening to $\mathcal{O}(5\%)$ 
at the phenomenologically irrelevant smallest values of $g_i$. This applies 
to the subtracted and tuned scheme results, for which \eq\eqref{fitdef}
are approximate forms of the numerical results. In the strict LO case 
it is on the other hand an exact result, up to the numerical accuracy of the $k_i$
coefficients.

We observe that the leading-logarithmic term and the cubic term in $g_i$ are fitted by the same respective 
values of $c_i$ and $\delta_i$ for the two non-abelian groups for each given scheme.
This is not surprising: as we mentioned before, in this normalization the gauge-boson exchange term
is the same for the two groups. As it is only this term that the three schemes treat differently, the observed 
scheme-by-scheme agreement of $c_2$ and $c_3$, $\delta_2$ and $\delta_3$ is then expected.

%
\begin{figure}[t]
\centerline{
  \includegraphics[width=0.74\linewidth]{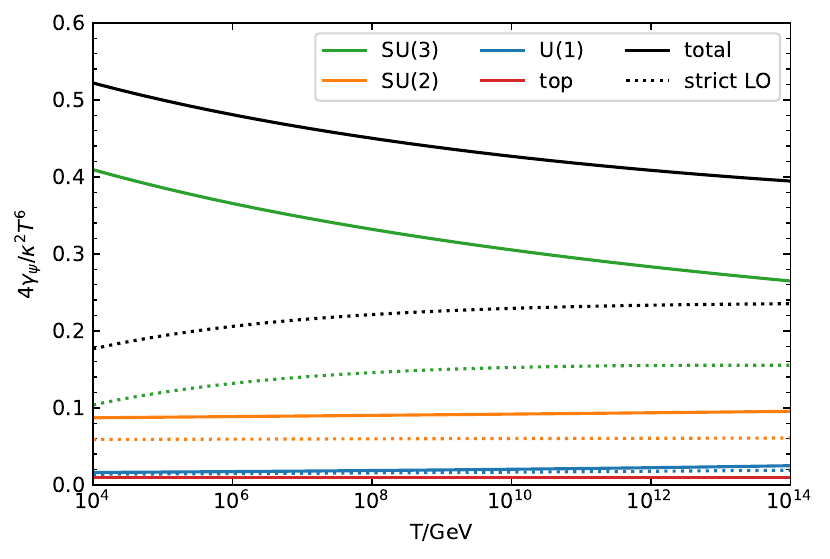}
}
\vspace{-3mm}
\caption[a]{\small
  The integrated rate $\gamma_\psi$ and its gauge and 
  top Yukawa components as a function of the temperature. 
  Solid lines arise from the tuned scheme, dashed ones from 
  the strict LO scheme.
}
\label{fig:gammaT}
\end{figure}
%

In \fig\ref{fig:gammaT} we plot the integrated rate 
$
  \gamma_{\psi}
  \equiv 
  2\int {\rm d}^3\vec{k}/(2\pi)^3 \, \Gamma_{\psi}(k)\nF^{ }(k)
$
as a function of the temperature. 
The factor of four in our $4/(\kappa^2T^6)$ normalisation 
has been chosen so as to reproduce exactly that of 
\fig12 in Ref.~\cite{Rychkov:2007uq}, 
\fig2 of Ref.~\cite{Eberl:2020fml} and 
\fig5 of Ref.~\cite{Eberl:2024pxr}. 
The temperature dependence arises 
from picking a prescription for the running of 
$g_i$ and $|h^{ }_t|$ as a function of $T$.
We follow the prescription used in the literature and 
adopt a constant $|h^{ }_t|=0.7$ and one-loop running for the $g_i$ with 
the aforementioned unified GUT value, i.e.
\begin{equation}
  \label{running}
  g_i^2(T)
  \; = \;
  \frac{g_i^2(M_\mathrm{GUT})}{
    \displaystyle 1-\frac{b_i}{8\pi^2} \, 
    g_i^2(M_\mathrm{GUT}) \,
    \ln\frac{\pi T}{M_\mathrm{GUT}}
    }
    \; ;
  \quad b_i=(11,1,-3)
  \, , \ 
  g_i^2(M_\mathrm{GUT})=\left(\frac{\pi}{10},\frac\pi6,\frac\pi6\right)\,.
\end{equation}
Unlike in previous analyses,
and in keeping with common practice in Thermal Field Theory (see e.g.~\cite{Laine:2005ai})
we have set 
the renormalization scale to $\bar\mu=\pi T$ rather than $\bar\mu=T$. 
The figure clearly shows how the spread between the total rate in the tuned and strict LO schemes,
denoted by the solid and dashed black lines respectively, grows from an $\mathcal{O}(50\%)$ effect
at the highest considered temperature to a factor of 3 on the opposite end. As expected,
given the larger values of $g_3$ and $d_3$, the SU(3) contribution dominates the rate and its 
uncertainty.

%
\begin{figure}[t]
\centerline{
  \includegraphics[width=0.9\linewidth]{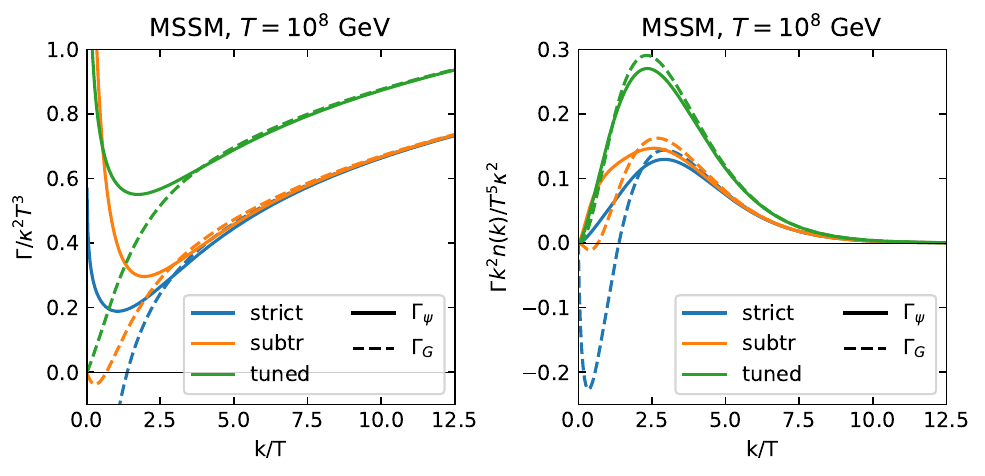}
}
\vspace{-3mm}
\caption[a]{\small
  The gravitino ($\psi$ component, solid lines) 
  and graviton rates ($G$, dashed lines)
  in the MSSM at $T=10^8$ GeV. On the left we plot the rates
  normalized by $\kappa^2T^3$ in the three schemes. On the 
  right we further multiply by $(k/T)^2 n(k)$, 
  where $n(k)$ is $\nF(k)$ and $\nB(k)$
  for gravitinos and gravitons respectively, 
  so that the area under each 
  curve is then proportional to the corresponding $F_i\,$.
}
\label{fig:susy}
\end{figure}
%

In \fig\ref{fig:susy} we compare the gravitino rate $\Gamma_\psi$ with 
its graviton counterpart, following the discussion in sec.~\ref{sec_gw_susy}.
As expected from the SUSY-based theoretical arguments~\cite{Caron-Huot:2008vbk}, the two
rates agree, scheme by scheme, in the $k\gg T$ range where the SUSY-breaking effects of quantum
statistics become negligible.  Conversely,
the IR is where they are the largest: the left panel of \fig\ref{fig:susy} 
shows that $\Gamma_\psi(k\ll T)>\Gamma_G(k\ll T)$. 
In principle this hierarchy becomes less severe 
in the right panel of \fig\ref{fig:susy}, where $\Gamma$ is multiplied
by $k^2$ times the statistical factor, since $\nB(k\ll T)\approx T/k\gg \nF(k\ll T)\approx 1/2\,$.
However, the strict LO version of $\Gamma_G(k\ll T)$ extrapolates to negative values
at the chosen temperature; the thermal factor $\nB(k\ll T)\approx T/k\gg 1$ 
serves only to amplify this unphysical effect.

%
\section{Conclusions}
\label{sec:concl}

In this paper we have employed novel automation techniques, 
facilitated by the \acrs{} package~\cite{autotherm}, 
for thermal gravitino and axino rates.  
\acrs{} is released as a public code~\cite{autosite}, 
ready to be used in cosmological studies. 
Through this tool, 
the tedious and error-prone determination of the matrix elements squared
can be carried out effortlessly; in addition, 
its main strength lies in the built-in, automated
implementations of Hard Thermal Loop resummation, 
which minimize the risk of methodological or 
computational errors  and enable comparisons between different 
leading-order-equivalent schemes.

Sections~\ref{sec:thprod} and \ref{sec:auto} 
are intended to provide a step-by-step, pedagogical 
overview of the calculation
starting from the field-theoretical definition 
of the production rate and 
its relation to the supercurrent. 
The implementation within \acrs{} and the 
workflow stages are described,
namely a determination of the Feynman rules, 
the evaluation of $\twotwo$ matrix elements squared 
with one final-state gravitino and the handling of HTL resummation. 
Our analytical results for the matrix elements squared are 
in agreement with earlier 
results~\cite{Bolz:2000fu,Pradler:2006qh,Pradler:2006tpx,Pradler:2006hh,Rychkov:2007uq}.

In sec.~\ref{sec:res} we then present our numerical results. 
We provide a slight update to the results 
of~\cite{Rychkov:2007uq} for the 
top-Yukawa contribution in Eq.~\eqref{topcontrib}  
and footnote~\ref{foot_yukawa}.  Furthermore,
our automated implementation of the \emph{strict leading order} 
rate---whereby the soft gauge-boson exchange
is separated by a cutoff from the remainder of the calculation 
and treated with HTL resummation---agrees 
with the classic results in 
Refs.~\cite{Bolz:2000fu,Pradler:2006qh,Pradler:2006tpx,Pradler:2006hh}. 
As is well known, this procedure~\cite{Braaten:1991dd} relies on the 
existence of a scale separation between the \emph{hard scale}
$k\gtrsim T$, where $k$ is the energy of the gravitino, 
and the \emph{soft scale} $g_iT$, with $g_i$ the gauge coupling;
its extrapolation to $k\lesssim g_i T$ is badly behaved, 
giving rise to unphysical, negative rates. 

We have identified pathologies in existing gauge-dependent 
resummation schemes~\cite{Rychkov:2007uq}
that were introduced to mend the issue of negativity by constructing a 
positive-definite rate at all $k$. 
As we show in appendix~\ref{app_slava}, such gauge-dependent 
resummations are especially incompatible with non-abelian gauge
theories, where ultrasoft magnetostatic  gauge bosons with energy 
and momentum $\vert q^0\vert \ll q\sim g_i^2 T$ are 
inherently non-perturbative~\cite{Linde:1980ts}.
The method of~\cite{Rychkov:2007uq} includes the (gauge-fixed) 
one-loop contribution to the magnetostatic self-energy;
the breakdown of perturbation theory for this quantity is reflected 
in their calculation by the emergence of a non-integrable 
divergence for $\vert q^0\vert \ll q\sim g_i^2 T$ which 
has apparently been missed in previous numerical 
implementations~\cite{Rychkov:2007uq,Eberl:2020fml,Eberl:2024pxr}.

In addition to the strict LO, we provide the 
\emph{subtracted} and \emph{tuned} schemes.
The former amounts to replacing 
$\ln(T/(g_iT)) \to 1/2 \ln(1+(T/(g_iT))^2)$ in the logarithmic part 
of the strict LO rate. 
This can be motivated from HTL sum 
rules~\cite{Caron-Huot:2008dyw,Ghiglieri:2020mhm}, 
but it is not sufficient to fully eliminate negative extrapolations. 
Our \emph{tuned} scheme, on the other hand, is  positive-definite by 
construction; it is free of any gauge choice pathology and it 
agrees with the strict leading order at asymptotically small couplings. 
We are thus able to quantify the ``theory uncertainty'', 
by comparing these three LO-equivalent schemes. 
In particular, we provide new fits for the fully integrated 
rate in table~\ref{tab:fitsfinal}, 
and the spread between schemes can be seen in 
figure~\ref{fig:gammaT}. It
ranges from $\mathcal{O}(50\%)$ in the vicinity of the GUT-scale
temperatures to factors of 3 at $T=10$ TeV. These are our main results.

Assuming that reheating rapidly establishes 
a radiation-dominated plasma with a maximal temperature 
of $T_\mathrm{RH}\,$, 
the integrated gravitino production rate 
(accumulated over the subsequent history)
determines the 
primordial yield $Y^{ }_{\tilde G} \equiv n^{ }_{\tilde G}/s\,$.
At late times,  $T \ll T_\mathrm{RH}$, 
this yield is approximately proportional to 
$\gamma^{ }_{3/2}(T_\mathrm{RH})\,$, although the coefficient depends 
on reheating dynamics as well as the thermal content.  
Ultimately, given $Y_{\tilde G}$ and the gravitino mass,  
one obtains its present day energy density  
and hence its possible contribution to the observed DM abundance
or to other facets of the ``gravitino problem''.

%
\section*{Acknowledgements}

We thank Helmut Eberl, Ioannis Gialamas, Mikko Laine, Slava Rychkov, Vassilis Spanos, Alessandro Strumia,  for useful discussions.
J.G.\ and G.J.\ are partly funded by the Agence
Nationale de la Recherche under grant ANR-22-CE31-0018 (AUTOTHERM).

%
\appendix
\renewcommand{\thesection}{\Alph{section}} 
\renewcommand{\thesubsection}{\Alph{section}.\arabic{subsection}}
\renewcommand{\theequation}{\Alph{section}.\arabic{equation}}

%
\section{Gauge-dependent resummations in the literature}
\label{app_slava}

%
\begin{figure}[t]
\centerline{
  \includegraphics[width=0.7\linewidth]{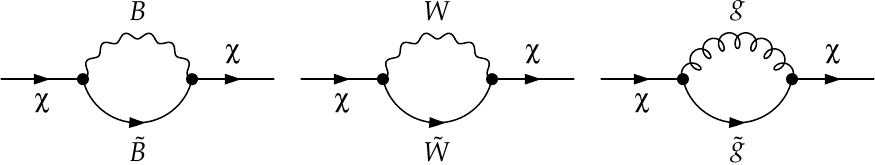}
}
\vspace{-1mm}
\caption[a]{\small
  The gravitino self-energy at one loop in the MSSM. 
  As the figure is generated by \textsc{FeynArts} from our model file, 
  only the $\chi$ component is displayed. 
  Internal lines are to be considered as resummed, 
  as discussed in the main text. 
  Here $B\,$, $W\,$, and $g$ are the usual gauge bosons, 
  while $\tilde{B}\,$, $\tilde{W}\,$, and $\tilde{g}$ are 
  their supersymmetric counterparts. 
}
\label{fig:selfdiags}
\end{figure}
%

In this section we summarize the detailed analysis 
of the gauge-dependent resummations of Ref.~\cite{Rychkov:2007uq} 
carried out by two of us for the axion  (cf. Ref.~\cite{Bouzoud:2024bom} for more technical details) 
and extend it to the case of the gravitino. 

In a nutshell, the method proposed in Ref.~\cite{Rychkov:2007uq} relies on
considering the one-loop diagrams in \fig\ref{fig:selfdiags}.
The relation we have used to obtain \eq\eqref{rate_gen} also connects 
the rate to the imaginary part of 
the gravitino self-energy~\cite{Bodeker:2015exa},  which 
is in turn related to the cuts of the corresponding diagrams. 
If the gauge boson and gaugino propagators appearing in \fig\ref{fig:selfdiags} 
were bare ones, 
the cuts would vanish kinematically for massless gravitinos and gauginos. 
If, on the other hand, they are resummed, non-trivial cuts 
can arise. To see this in more detail, let us start by 
amending \eq(4.7) of Ref.~\cite{Rychkov:2007uq} so as to retain the correct number of integrals, 
as in Ref.~\cite{Eberl:2024pxr}. 
We then have, for a each gauge group $i$ with coupling $g_i$ 
\begin{align}
    \Gamma_\psi(k)\bigg\vert^\text{\cite[Eq.(4.7)]{Rychkov:2007uq}}_\text{gauge}
    =&
   \frac{d_i \kappa^2 }{4(4\pi)^3 k^2}
    \int_{-\infty}^\infty \! \mathrm{d}q^0
    \int_0^\infty \! \mathrm{d}q
    \int_{\vert q-k\vert}^{q+k} \! \mathrm{d}p 
   \,  q\, 
   \big[1+\nB(q^0)-\nF(k-q^0)\big]\nonumber \\[2mm]
   \times &
   \bigg\{
   \pm\rho^{FG}_{L}( \Q )
   \rho^{FG}_\pm( \P ) \, (k\pm p)^2 \, 
   \big[ q^2 - (k\mp p)^2 \big]
   \nonumber \\[2mm]
   & \hspace{-1.5cm}
   \pm\rho^{FG}_T(\Q)
   \rho^{FG}_\pm(\P) \,
   \big[ (k\pm p)^2 - q^2 \big] \, 
   \bigg[
     \left(1+\frac{q_0^2}{q^2}\right)
      \big(q^2+(k\mp p)^2\big)
      -4q^0(k\mp p)
   \bigg]
   \bigg\}
   \, ,
   \label{startsalvio2} 
\end{align}
where we have used the identity 
$\nB(q^0)\nF(k{-}q^0)/\nF(k)=1+\nB(q^0)-\nF(k{-}q^0)$ 
and set $p^0=k{-}q^0$. 
This gauge part is to be complemented by the  
residual $2\leftrightarrow 2$
component to assemble the LO contribution, 
such as those arising from Yukawa couplings, 
that are not included in \eq\eqref{startsalvio2}; 
we refer to Refs.~\cite{Rychkov:2007uq,Eberl:2024pxr} 
for further details.
Above, $\rho^{FG}_{L,T}(\Q)\equiv G^R_{L,T}(\Q)-G^A_{L,T}(\Q)$ are the Feynman-gauge 
resummed vector boson
spectral densities and $\rho^{FG}_\pm(\P)$ the resummed gaugino ones. 
$L$ and $T$ denote modes 
that are respectively longitudinal and transverse to the three-momentum $q^i$, whereas 
$\pm$ denote the $\pm 1$ value of the helicity-to-chirality ratio. 
A sum over these two signs is implied in \eq\eqref{startsalvio2}. 
In vacuum, 
propagating massless
fermionic particles (antiparticles) have a positive (negative) ratio and propagating, 
massless gauge bosons are spatially transverse.

Let us now discuss the specifics of resummation. 
Explicitly, the Feynman-gauge spectral functions 
going into \eq\eqref{startsalvio2} read\footnote{%
\label{foot_widths}
  Here we assumed that the imaginary parts 
  vanish above the light cone, thus 
  the pole widths are neglected. 
}
\begin{eqnarray}
  \label{spf_T}
  \rho^{FG}_T(\Q)
  & = &
  \left\{
  \begin{array}{lc}
  \displaystyle
       \frac{
         -2\,\mathrm{Im}\,\Pi_R^T(\Q)
        }{
          \bigl[\Q^2 
          -\mathrm{Re}\,\Pi_R^T(\Q)\bigr]^2+
          \bigl[\mathrm{Im}\,\Pi_R^T(\Q)\bigr]^2
        }
       & \text{for}\ \ |q^0_{ }| < q 
       \; , \\[6mm]
       \displaystyle
       2 \pi \, \sign(q^0_{ }) \, \delta \Big( \Q^2_{ } - \mathrm{Re}\,\Pi_R^T(\Q) \Big)
       & \text{for}\ \ |q^0_{ }| > q 
       \; ,
  \end{array}
  \right.
  \\[3mm]
  \label{spf_L}
  \rho^{FG}_L(\Q)
  & = &
  \left\{
  \begin{array}{lc}
  \displaystyle
       \frac{
         -2\,\mathrm{Im}\,\Pi_R^L(\Q)
        }{
          \bigl[q_{ }^2 
          +\mathrm{Re}\,\Pi_R^L(\Q)\bigr]^2+
          \bigl[\mathrm{Im}\,\Pi_R^L(\Q)\bigr]^2
        }
       & \text{for}\ \ |q^0_{ }| < q 
       \; , \\[6mm]
       \displaystyle
       2 \pi \, \sign(q^0_{ }) \, \delta \Big( q^2_{ } + \mathrm{Re}\,\Pi_R^L(Q) \Big)
       & \text{for}\ \ |q^0_{ }| > q 
       \; ,
  \end{array}
  \right.
  \\[3mm]
  \label{spf_pm}
  \rho^{FG}_\pm(\P)
  & = &
  \left\{
  \begin{array}{lc}
  \displaystyle
       \frac{
         -2\,\mathrm{Im}\, \Sigma^{ }_\pm(\P)
        }{
          \bigl[ p_{ }^0 \mp p - \mathrm{Re}\,\Sigma^{ }_\pm(\P)\bigr]^2+\bigl[\mathrm{Im}\,\Sigma^{ }_\pm(\P)\bigr]^2
        }
       & \text{for}\ \ |p^0_{ }| < p 
       \; , \\[6mm]
       \displaystyle
       2 \pi \, \sign(p^0_{ }) \, \delta \big( p_{ }^0 \mp p- \mathrm{Re}\,\Sigma^{ }_\pm(\P) \big)
       & \text{for}\ \ |p^0_{ }| > p 
       \; ,
  \end{array}
  \right.
\end{eqnarray}
where the particular self-energies are given in 
Feynman gauge as 
$
  \Sigma(\P) 
  \equiv 
  \frac{ \gamma^{ }_0 + \hat p \cdot \vec{\gamma} }{2} 
  \Sigma^{ }_+ (\P)
  +
  \frac{ \gamma^{ }_0 - \hat p \cdot \vec{\gamma} }{2} 
  \Sigma^{ }_- (\P)
$ 
and 
$
  \Pi^{ }_{\mu \nu}(\Q)
  \equiv
  \mathbb{P}^T_{\mu\nu} \Pi^{T}_{ }(\Q)
  -
  \frac{\Q^2}{q^2}
  \mathbb{P}^L_{\mu\nu} \Pi^{L}_{ }(\Q)
$ 
using the projectors 
$
  \mathbb{P}^T_{\mu\nu}
  =
  g^{ }_{\mu i} g^{ }_{\nu j} ( \delta^{ }_{ij} - 
  \frac{ q^{ }_i q^{ }_j }{ q_{ }^2 } )
$
and 
$
  \mathbb{P}^L_{\mu\nu}
  = 
  g^{ }_{\mu\nu} - \frac{ \Q^{ }_\mu \Q^{ }_\nu }{ \Q_{ }^2 } - 
  \mathbb{P}^T_{\mu\nu}
  \, 
$.
In Ref.~\cite{Rychkov:2007uq}  
the full one-loop Feynman-gauge retarded self-energy
is resummed for gauge bosons and gauginos alike when their
momenta $\Q$ or $\K{-}\Q$ are space-like. 
Conversely, for time- and light-like
momenta they resum the self-energy in the HTL approximation, 
thus giving rise to zero-width collective
excitations stemming from thermally modified 
dispersion relations~\cite{Weldon:1982aq,Weldon:1982bn}. 
References~\cite{Eberl:2020fml,Eberl:2024pxr}
use instead the full one-loop (see e.g. Ref.~\cite{Peshier:1998dy}) 
Feynman-gauge spectral function in all kinematical regimes.  
In either case, this procedure is claimed to reduce to the strict LO rate at small $g_i$
and to have a residual gauge dependence of relative order $g_i^2$.
As in the case of the axion, these claims do not hold, as we now illustrate in some detail.

\begin{figure}[t]
\centerline{
  \includegraphics[width=0.33\linewidth]{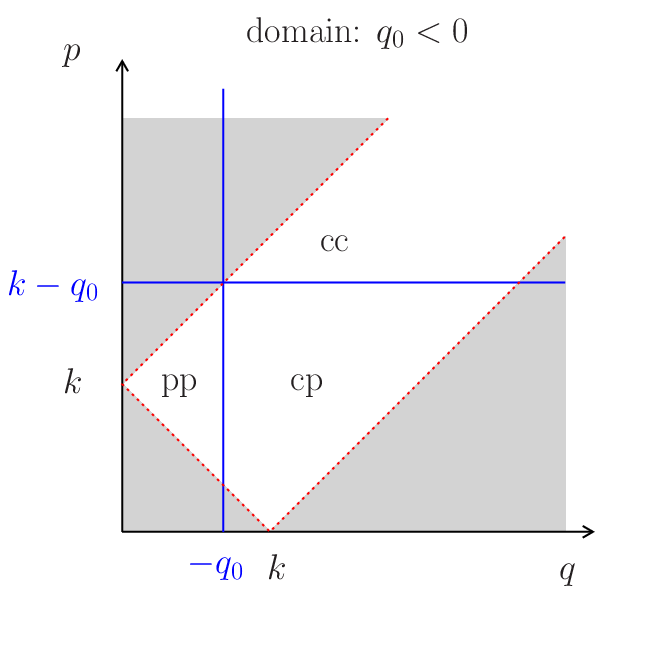}
  \includegraphics[width=0.33\linewidth]{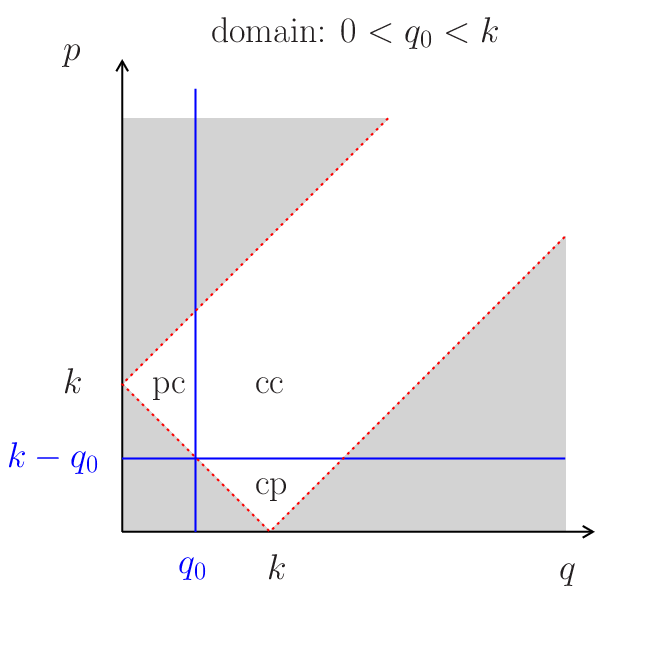}
  \includegraphics[width=0.33\linewidth]{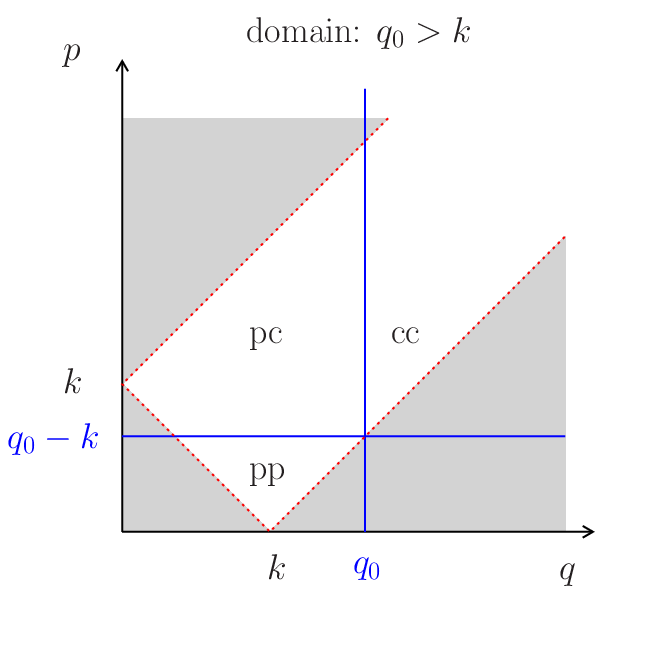}
}
\vspace{-5mm}
\caption[a]{\small 
  Domains of the $p$ and $q$ integration in \eq\eqref{startsalvio2}, 
  where grey regions are inaccessible. 
  The solid blue lines separate regions with different natures 
  of the gauge boson and gaugino spectral functions (whether it describes 
  Landau damping or a pole contribution). 
  The arrangement of these regions, and realised configurations, 
  depends on whether $q_{ }^0$ is on the interval 
  $(-\infty,0]\,$, $[0,k]\,$, or $[k,\infty)\,$, 
  as indicated above each plot.  
}
\label{fig:poles_and_cuts}
\end{figure}

To be clear, let us first assess the nature of the 
spectral functions in different domains of integration 
according to \eq\eqref{startsalvio2}. As mentioned, 
the four-momenta $\Q$ and $\P$ are associated to the gauge boson 
and gaugino respectively. We keep this ordering when referring 
to whether the corresponding spectral function has a 
{\em cut} (space-like) or a {\em pole} (time-like). 
The diagrams in \fig\ref{fig:selfdiags} 
give rise to four different type of contributions.
Recalling that $p_{ }^0 = k - q_{ }^0\,$, 
the explicit breakdown is 
\begin{eqnarray}
  | q_{ }^0 | < q 
  \hspace{6mm} \text{and} \hspace{6mm}
  |k - q_{ }^0| < p 
  \hspace{4mm}
  & \Rightarrow & 
  \hspace{4mm}
  \text{``cut-cut'' \ \ \, \, (cc)} 
  \; , \\[2mm]
  | q_{ }^0 | < q 
  \hspace{6mm} \text{and} \hspace{6mm}
  |k - q_{ }^0| > p 
  \hspace{4mm}
  & \Rightarrow & 
  \hspace{4mm}
  \text{``cut-pole'' \ \ \, (cp)} 
  \; , \\[2mm]
  | q_{ }^0 | > q 
  \hspace{6mm} \text{and} \hspace{6mm}
  |k - q_{ }^0| < p 
  \hspace{4mm}
  & \Rightarrow & 
  \hspace{4mm}
  \text{``pole-cut'' \ \ \, (pc)} 
  \; , \\[2mm]
    | q_{ }^0 | > q 
  \hspace{6mm} \text{and} \hspace{6mm}
  |k - q_{ }^0| > p 
  \hspace{4mm}
  & \Rightarrow & 
  \hspace{4mm}
  \text{``pole-pole'' \ \,\! (pp)} 
  \; .
\end{eqnarray}
Whether certain conditions are actually realised in the $(p,q)$-plane 
depends on the value of $q_{ }^0$ 
(i.e., the outer integration in \eq\eqref{startsalvio2}). 
A simple case-by-case geometric exercise leads to the disjoint 
regions depicted in Fig.~\ref{fig:poles_and_cuts}. 
We now proceed to examine each case.

%
%

First, the pole-pole contribution occurs 
where both the gauge boson and the gaugino are on their time-like 
plasmon (or plasmino) poles. 
The scenario with $q_{ }^0 < 0$ (and thus $p_{ }^0 > k$) in 
Fig.~\ref{fig:poles_and_cuts} 
evidently corresponds to processes like
$\tilde{g} \leftrightarrow g\,\psi$, 
where $g$ and $\tilde{g}$ are on-shell plasmon/plasmino excitations. 
Similarly, the scenario with $q_{ }^0 > k$ (and thus $p_{ }^0 < 0$) 
corresponds to processes like $g \leftrightarrow \tilde{g}\,\psi\,$.\footnote{%
  In order to see this, 
  it is useful to reinstate 
  $\int {\rm d}p^{ }_0 \delta(k-p^{ }_0-q^{ }_0)$ 
  in \eqref{startsalvio2}, 
  and subsequently carrying out the $p^0_{ }$ 
  and $q^0_{ }$ integrations using the 
  Dirac-$\delta$ parts from \eqs\eqref{spf_T}--\eqref{spf_pm}. 
  That leaves the $p,q$ integrals 
  in the two regions from \fig\ref{fig:poles_and_cuts} subject to  
  $k = q^{T,L}_0(q) - p^{\pm}_0(p)$ 
  or 
  $k = - q^{T,L}_0(q) + p^{\pm}_0(p)\,$. 
  We thus see that the emission of $\psi$ involves a 
  transition between branches of the dispersion relation 
  for the gauge boson and its supersymmetric gaugino partner, 
  subject to energy and momentum conservation. 
} 
At first Ref.~\cite{Rychkov:2007uq} argues that the thermal masses of the
gauge-boson and gaugino components of each gauge multiplet are different, 
which would then imply a 
significant contribution from these processes. However, as they correctly argue later on,
``thermal masses'' are momentum-dependent. 
While in the zero-momentum plasmon limit they are indeed different, 
the {\em asymptotic masses} $m_{\infty i}$ 
are the same for members of 
the same supersymmetric (gauge or chiral) multiplet~\cite{Caron-Huot:2008vbk}, 
implying
\begin{equation}
    q_0^{T}\big(q \gg g^{ }_i T\big) 
    \; \simeq \;
    \sqrt{ m^2_{\infty i} + q^2_{ }}
    \ , \qquad 
    p_0^{+}\big(p \gg g^{ }_i T\big) 
    \; \simeq \; 
    \sqrt{ m^2_{\infty i} + p^2_{ }}
    \; .
\end{equation}
(The $L$-mode and plasmino approach the light cone, although with
exponentially suppressed residues.) 
This makes it clear that if both $g$ and $\tilde g$ are 
hard, there is no available phase space 
for $1 \to 2$ gravitino emission when 
both $\P$ and $\Q$ are hard.\footnote{%
  More generally, these processes are proportional 
  to the (asymptotic) mass difference 
  which would involve subleading SUSY-breaking terms. 
  We do not consider those here. 
}
When both $\P$ and $\Q$ are soft, 
and thus $k \ll T$, a careful analysis shows that
this region does not contribute to leading order~\cite{Bouzoud:2024bom}. 
A small corner of the phase space remains 
where either i) $\Q$ is soft and $p \sim k$ is hard 
(leftmost panel in \fig\ref{fig:poles_and_cuts}), 
or ii) $\P$ is soft and $q \sim k$ is hard 
(rightmost panel in \fig\ref{fig:poles_and_cuts}).
As eventually stated in 
Ref.~\cite[footnote 4]{Rychkov:2007uq}, 
this pole-pole contribution is numerically small.

The next types of contribution we discuss are 
the cut-pole and pole-cut ones. 
In the former the gauge boson in
its spacelike Landau cut, whereas the gaugino is on its plasmon or plasmino pole; the opposite happens in the latter case. 
We note that the last type, namely the cut-cut one,
involves both the gauge boson and the gaugino being space-like. 
As shown in Ref.~\cite{Bouzoud:2024bom} the cut-cut contribution too is subdominant---though potentially
also affected by the main issue plaguing the 
cut-pole part of \eq\eqref{startsalvio2}.
Let us then analyze the cut-pole part carefully. 

At small $g_i$ and for $k\gtrsim T$ we can  
neglect the exponentially suppressed ``hole'' or ``plasmino'' modes,
which at positive (negative) frequency appear in $\rho_-$ ($\rho_+$).
We can then approximate the gaugino plasmon modes by its bare one 
in \eq\eqref{spf_pm}, i.e.
\begin{equation}
  \rho^{\text{pole}}_\pm
  \big( p \gg g^{ }_i T \big)
  \; \approx \; 
  2\pi\delta\big(
    \underbrace{(k-q_{ }^0)}_{p_{ }^0} \mp p
  \big)
  \; .
  \label{bareplasmon}
\end{equation}
From Fig.~\ref{fig:poles_and_cuts}, 
the cut-pole can only occur when $k-q_{ }^0 > 0\,$, 
thus restricting phase space to the $+$ part. 
Returning to \eq\eqref{startsalvio2}, 
and using \eqref{bareplasmon} to carry out the 
$p$-integral, 
\begin{align}
  \Gamma_\psi(k)\bigg\vert^\text{\cite{Rychkov:2007uq} cp}_\text{gauge}
  \; = \; &
  \frac{2\pi  d_i \kappa^2}{4(4\pi)^3 k^2}
  \int_{-\infty}^k \!\mathrm{d}q^0
  \int_{\vert q^0\vert}^{2k-q^0} \!\mathrm{d}q\, q\, 
  \big[1+\nB(q^0)-\nF(k-q^0)\big](q^2-q_0^2)\nonumber \\[2mm]
  \; \times \;& 
  \bigg\{
  \rho^{FG}_L(\Q)\big(2k-q^0\big)^2-
  \rho^{FG}_T(\Q) \big[(2k-q^0)^2-q^2\big]\frac{\Q^2}{q^2}
  \bigg\}
  \, ,
  \label{salviosanity2} 
\end{align}
where cp stands for cut-pole.
We now  turn to the IR sector of this  expression, where the intermediate gauge boson is soft. 
For $q^{ }_0,q \sim g^{ }_i T\ll k$ we find
\begin{align}
  \Gamma_\psi(k)\bigg\vert^\text{\cite{Rychkov:2007uq} cp soft}_\text{gauge}
  \; = \;&
  \frac{ d_i \kappa^2 }{2(4\pi)^2}
  \int_{-\infty}^k \!\mathrm{d}q^0
  \int_{\vert q^0\vert}^{2k-q^0} \!\mathrm{d}q \,q\, 
  \frac{T}{q^0}(q^2-q_0^2)
  \bigg[
    \rho^{FG}_L(\Q)-\rho^{FG}_T(\Q)\frac{\Q^2}{q^2}
  \bigg] 
  \, .
  \label{salviosanity3} 
\end{align}

As shown at length in Ref.~\cite{Bouzoud:2024bom}, a problem 
occurs because the Feynman-gauge one-loop transverse space-like spectral function \emph{does not} reduce 
to its HTL limit\footnote{\label{foot_htl}%
If one replaces $\rho^{FG}_{L,T}$ with their HTL counterparts $\rho^\mathrm{HTL}_{L,T}$ in 
\eq\eqref{salviosanity3}, one finds, using the light-cone analyticity
techniques of~\cite{Aurenche:2002pd,CaronHuot:2008ni}
and the variable changes discussed in~\cite{Ghiglieri:2020mhm,Bouzoud:2024bom}
\begin{align}
  \Gamma_\psi(k)\bigg\vert^\text{\cite{Rychkov:2007uq} cp HTL}_\text{gauge}
  \; = \;&
 \frac{ d_i\kappa^{2}}{32 \pi} T \mDi^2 \ln\left(1+\frac{4 k^{2}}{\mDi^2}\right)
  \, ,
  \label{salviosanityHTL} 
\end{align}
which is precisely the logarithmic term in \eq\eqref{subtrlog}.} 
for $\vert q^0\vert \ll q\sim g_i^2 T$, with pathological consequences 
in the non-abelian case. 
Explicitly, the transverse self-energy 
in Feynman gauge reads\footnote{%
  The following is a slightly more compact way of expressing 
  Eqs.~(B.3) from Ref.~\cite{Rychkov:2007uq}. 
} (here $\bar \mu$ is the renormalisation scale)
\begin{eqnarray}
    \label{eq:full_piT_FG}
    \Pi^{T}_{R} (q^0 , q) 
    & \stackrel{|q^0_{ }| < q}{=} &
    \frac{g^2}{2} \left\{ 
    \frac{-\Q^2}{24 \pi^2} \big(5 \ncol - 2 \nferm - \nscal \big) \ln \frac{- \Q^2}{\bar \mu^2}
    \right. 
    \\[2mm]
    & + & 
    \frac{T^2}{6 q^2}
    \big( \nscal + \tfrac12 \nferm + \ncol \big) 
    \left( q^2_0 + q^2 - \frac{q^{ }_0 \Q^2}{q} 
    \ln \frac{q^{ }_+}{q^{ }_-} \right)
    \nonumber\\[2mm]
    & + &
    \left.
    \frac{\Q^2}{2 \pi^2 q^3} \Big[ 
    (\ncol + \nscal) \, {\cal J}^{ }_{\rm B} + 
    \nferm \, {\cal J}^{ }_{\rm F}
    + q^2
    \big( \ncol \, {\cal I}^{ }_{\rm B} 
    + \tfrac12 \nferm \, {\cal I}^{ }_{\rm F}
    \big)
    \Big]
    \right\}
    \hspace{12mm}
    \; ,
    \nonumber
\end{eqnarray}
where $q^{ }_\pm \equiv \tfrac12 (q^{ }_0 \pm q)$ and the ``master integrals'' are defined by (with $\sigma=\{ \text{\footnotesize B, F} \}$)
\begin{eqnarray}
    \label{eq:J_I_defs}
    {\cal J}^{ }_\sigma  (q^0 , q)
    & \equiv & 
    \int^\infty_0 {\rm d}p
    \left\{
    (p+q^{ }_+)(p+q^{ }_-) \ln \frac{p+q^{ }_+}{p+q^{ }_-}
    -
    (p-q^{ }_+)(p-q^{ }_-) \ln \frac{p-q^{ }_+}{p-q^{ }_-}
    \right\}
    n^{ }_\sigma (p) 
    \; ,
    \nonumber \\[2mm]
    {\cal I}^{ }_\sigma  (q^0 , q)
    & \equiv & 
    \frac{ \partial^2 {\cal J}^{ }_\sigma }{ \partial q^{ }_+ \partial q^{ }_-}
    \; .
\end{eqnarray}
For the real part, ${\cal J}_\sigma$ and ${\cal I}_\sigma$ must be evaluated numerically, but their imaginary parts are 
expressible in terms of polylogarithms.\footnote{\label{foot_poly}%
  For the imaginary part of ${\cal J}^{ }_\sigma  (q^0 , q)$ we find, 
  in the space-like domain
  \begin{align}
    \mathrm{Im}\,{\cal J}^{ }_\sigma  (\vert q^0\vert< q)=&
    -(1+\delta_{\sigma\mathrm{B}})\frac{\pi^3 T^2 q^0}{6}+
    (-1)^{\delta_{\sigma\mathrm{F}}}\pi q T^2\bigg[\mathrm{Li}_2\left((-1)^{\delta_{\sigma\mathrm{F}}}e^{q_-/T}\right)
    -\mathrm{Li}_2\left((-1)^{\delta_{\sigma\mathrm{F}}}e^{-q_+/T}\right)\bigg]\nonumber\\
    +&(-1)^{\delta_{\sigma\mathrm{F}}}2\pi  T^3\bigg[\mathrm{Li}_3\left((-1)^{\delta_{\sigma\mathrm{F}}}e^{q_-/T}\right)
    -\mathrm{Li}_3\left((-1)^{\delta_{\sigma\mathrm{F}}}e^{-q_+/T}\right)\bigg].
    \label{eq:imJpoly}
  \end{align}
} The relevant expansions of $\Pi^T_R$ needed below can be obtained from the corresponding expansions of these master functions, 
as shown in footnote~\ref{foot_expand}. 
In \eqs\nr{eq:full_piT_FG} and \nr{eq:J_I_defs}, the arguments of the logarithms should be understood with $q^0 \to q^0+ i 0^+$.

To understand where the issue arises, 
let us recall the structure of $\rho_R^T(\Q)$ 
from \eq\eqref{spf_T} for $|q^0_{ }| < q\,$. 
The subtlety arises in the limit
$\vert q^0\vert \ll q\sim g_i^2 T$. 
For such momentum arguments, the imaginary part reads
\begin{equation}
\label{impartTsmallfreq}
  \mathrm{Im}\,\Pi_R^T
  \big( q_0\to 0, q\lesssim g_iT \big)  
  \; = \;
  -\frac{\pi  \mDi^2 q^0}{4 q}-\frac{g_i^2 q^0T}{4\pi}(\ncol-\nscal)
  +\mathcal{O}\Big(g_i^2 q_0^3 T^2/q^2,g_i^2 q^0 q\Big)\,,
\end{equation}
where the first term on the r.h.s. is the HTL term and the second 
is its first correction. For non-abelian groups $\text{SU}(N)$, $\ncol=N$ is the degree of
the  gauge group; $N_i=0$ in the abelian case. $\nscal$ and, for later use,
$\nferm$, denote the number of scalars and fermions weighed by
Casimir squared or hypercharges; we  will discuss this momentarily.

%
\begin{figure}[t]
\centerline{
  \includegraphics[width=0.8\linewidth]{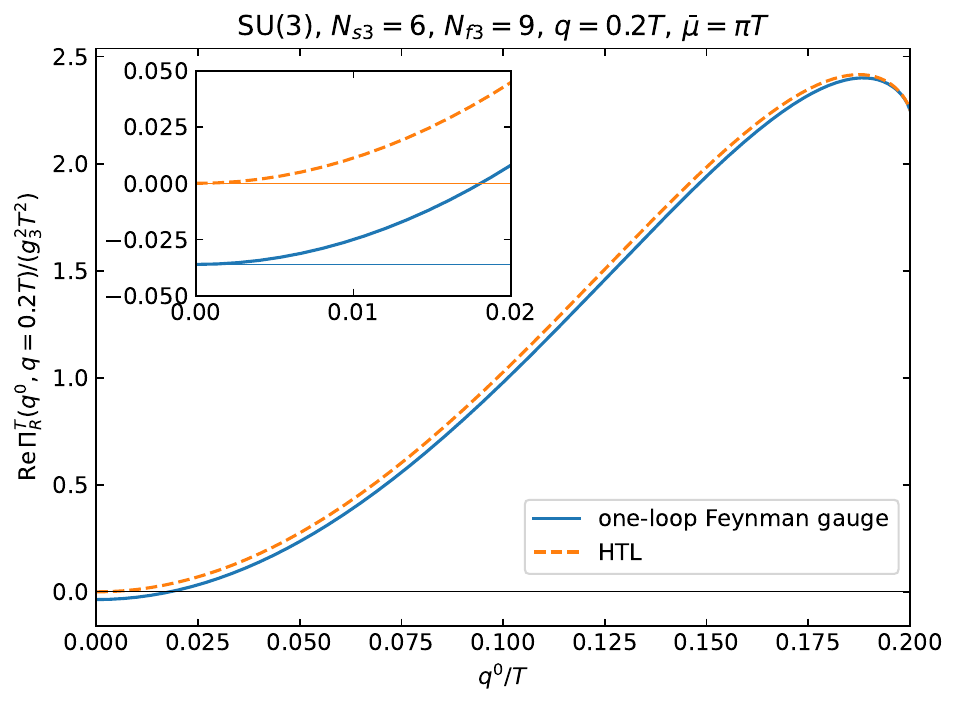}
}
\vspace{-3mm}
\caption[a]{\small
  The real part of the space-like retarded one-loop gluon 
  transverse self-energy, from Eq.~\eqref{eq:full_piT_FG},
  in the MSSM 
  at fixed momentum $q=0.2T$ as a function of the frequency. 
  The solid line 
  is the Feynman-gauge result, which includes the 
  (logarithmic part) of the vacuum 
  contribution at a renormalization scale $\bar\mu=\pi T$. 
  The dashed line is the 
  gauge-invariant, purely thermal HTL limit. 
  The inset displays a magnification of the 
  behaviour 
  at small frequency; 
  the thin horizontal lines are the zero-frequency limits.
  In Feynman gauge this limit is given 
  by Eq.~\eqref{trouble}; its HTL counterpart vanishes, 
  as explained in the main text.
}
\label{fig:rePiT}
\end{figure}
%

Hence $\mathrm{Im}\,\Pi_R^T(\Q)\propto q^0$, 
so that it vanishes at the denominator of 
\eq\eqref{spf_T}. 
This is expected on general grounds (causality), 
as ${\rm Im}\, \Pi_R(\Q)$ should be 
and odd function of $q_{ }^0$ whereas 
${\rm Re}\, \Pi_R(\Q)$ should be even. 
Thus the numerator in \eqref{spf_T} is linear in $q_{ }^0\,$, 
which is crucial in \eqs\eqref{salviosanity2} and \eqref{salviosanity3} 
because it compensates the would-be $T/q^{0}_{ }$ singularity 
from $\nB(q^{0}_{ })$ for small $q^0_{ }\,$. 
The important point is that,  though 
real and imaginary parts of the transverse HTL self-energy are of
order $g_i^2T^2$ by construction,  the real part vanishes for  $\vert q^0\vert \ll q\sim g_i^2 T$
and the subleading, gauge
dependent term takes over.
It reads,\footnote{\label{foot_expand}%
  The main ingredient is the expansion of the real part of 
  ${\cal J}^{ }_\sigma$,
  as given in \eq\eqref{eq:J_I_defs}, 
  for $\vert q^0\vert< q\ll T$. It can be achieved 
  by noting that, in this soft limit, ${\cal J}^{ }_\sigma$ 
  receives contributions 
  from the \emph{hard} scale $p\sim T\gg q$ and from 
  the \emph{soft} scale $p\sim q\ll T$. 
  In the former case one can expand the integrand 
  for small $q$ and $q_0$, while in the 
  latter case one needs to expand the statistical functions 
  as $\nB(p)=\frac{T}{p}-\frac12$ and 
  $\nF(p)=\frac12$. This yields
  \begin{align}
    \mathrm{Re}\,{\cal J}^{ }_\sigma  (\vert q^0\vert< q\ll T)
    = & \,
    (1+\delta_{\sigma\mathrm{B}}) \frac{\pi^2}{6} q T^2
    +
    \delta_{\sigma\mathrm{B}}
    \frac{\pi^2}{4}T\mathcal{Q}^2 
    \nonumber \\[1mm]
    - &
    (-1)^{\delta_{\sigma\mathrm{B}}}\frac{q^3}{12}
    \bigg[
      \ln\left(\frac{
        ((1{+}\delta_{\sigma\mathrm{B}})^2\pi T)^2
      }{
        e^{2\gamma_E}\vert \mathcal{Q}^2\vert
      }\right)
    +
    \left(\frac{q_0^2}{q^2}-3\right)\frac{q_0}{2q}
    \ln\left|\frac{q_+}{q_-}\right|
    +
    \frac{8}{3}
    -
    \frac{q_0^2}{q^2}
    \bigg]
    +
    \mathcal{O}\bigg(\frac{\mathcal{Q}^4}{T}\bigg).
  \end{align}
  The first, leading term is part of the HTL and thus arises 
  from the hard scale only. The second 
  term is purely bosonic and is the leading, 
  Bose-enhanced soft contribution from $\nB(p)\approx T/p$.
  Finally, the NNLO terms on the second line receive contributions 
  from both scales, as shown 
  by the logarithm of their ratio; an
  intermediate regulator is thus needed. 
  The soft contribution thereto
  arises from the $(-1)^{\delta_{\sigma\mathrm{B}}}/2$ term in the soft expansion of $n_\sigma(p)$.
} after expanding \eq\eqref{eq:full_piT_FG}, thus extending to scalar loops the QCD results in Refs.~\cite{Kalashnikov:1980tk,Kajantie:1982xx,Bouzoud:2024bom}
\begin{eqnarray}
    \mathrm{Re}\,\Pi_R^T \big(0,q\ll T\big)
    & = &
    \frac{ g_i^2  q T}{16}(\nscal-3\ncol)
    \label{trouble} \\[2mm]
    & + &
    \frac{2g_i^2 q^2}{(4\pi)^2} \bigg\{\left(\frac53 \ncol-\frac23  \nferm-\frac{\nscal}{3}\right) \ln\frac{4\pi T}{e^{\gamma_E}\bar\mu}
    \nonumber \\[2mm]
    & + &
    \frac{1}{9}\bigg[14\ncol-4\nscal
    +\nferm\bigg(12\ln2-5\bigg )\bigg]\bigg\}
    \; + \; 
    \mathcal{O}\left(\frac{g_i^2 q^3}{T}\right)\,,
    \nonumber
\end{eqnarray}
where $\gamma_E$ is the Euler--Mascheroni constant. The first, linear term in this $q/T$ expansion
arises from bosons only: it comes from an integration region where the loop momentum $p$ is soft too.
In the bosonic case this region receives a $T/p$ Bose enhancement.
Finite $q_{ }^0$ corrections to \eq\eqref{trouble} are necessarily quadratic 
for ${\rm Re}\,\Pi_R(\Q)$ to be an even function of $q^0_{ }$ as mentioned. 
The
HTL term is indeed quadratic in $q^0_ { }$ for $\vert q^0\vert \ll q$.

Let us consider the $i=3$ gluon case: in the MSSM
we have $N^{ }_{s\,3}=6$
(from the squarks)
and $N^{ }_{f\,3}=9$
(from $4 T^{ }_F N^{ }_g=6$ for the usual fundamental quarks and antiquarks, plus $T^{ }_A=3$ for the gluino), 
while in the SM 
$N_{s\,3}=0$ and $N_{f\,3}=6$. 
In \fig\ref{fig:rePiT} we plot the real part of the Feynman-gauge
and HTL one-loop transverse gluon self-energies in the space-like regime for a soft momentum $q\ll T$.
While the agreement between two is in general very good, the inset shows 
the negative dip of the Feynman-gauge expression at small frequencies,
emerging from the negative first term of \eq\eqref{trouble}.
We have then the same issue affecting the axion calculation: namely,
this negative first term brings 
a gauge-dependent pole into the $q$-integration domain 
when \eq\eqref{spf_T}
is plugged in \eq\eqref{salviosanity3}. The gauge dependence 
arises because the coefficient of the $\ncol$-proportional part of the linear-in-$q$
term in \eq\eqref{trouble} is gauge dependent and negative in all gauges~\cite{Kalashnikov:1980tk,Kajantie:1982xx}. In a generic covariant gauge parametrized by $\alpha$ it becomes
$-(8+(\alpha+1)^2)g_i^2  q T\ncol/64$. $\alpha=1$ denotes Feynman gauge.
This gauge-dependent pole in the zero-frequency transverse propagator for $q\sim g_3^2 T$
was first pointed out in the seminal paper by Linde~\cite{Linde:1980ts}, where the eponymous problem
was introduced. It signals the breakdown of perturbation theory at the non-perturbative ``ultra-soft'' 
scale $g_3^2T/\pi\,$; resumming the full one-loop transverse self-energy 
leads to this pathological pole in any gauge.\footnote{\label{foot_scalars}%
  In the case of QCD the only thermal, colored bosons are the gluons 
  themselves, so that the linear-in-$q$ term is 
  completely gauge-dependent and always negative. 
  In the present case there is in addition the
  gauge-invariant, positive squark contribution, 
  which can lift the pole in some gauges. 
  As we shall discuss, in the SU(2) case the squark, slepton and 
  Higgs contributions are large enough to remove 
  the pole in Feynman gauge.
} 

Following Ref.~\cite{Bouzoud:2024bom}, we examine the contribution 
of the pathological integration region to the gravitino rate,
finding for the SU(3) contribution
\begin{align}
   \label{startsalviosanityboomstart}
  \Gamma^{ }_\psi(k)
  \bigg\vert^\text{\cite{Rychkov:2007uq} cp div}_\text{SU(3)}
  \; \approx \; &
  \, \frac{ d^{ }_3 \kappa_{ }^2 T  m_\rmii{D3}^2 }{ 64\pi }
  \int_{\mathcal{O}(g_3^2 T)}^{ } \mathrm{d}q 
  \int_{-q_0^*}^{q_0^*} \, 
  \frac{
    \mathrm{d}q_{ }^0
  }{
    \big[ q(1+b^{ }_3)- a^{ }_3 \big]^2
    +
    \pi_{ }^2 m_\rmii{D3}^4 \frac{q_0^2}{16q^4}
    }
  \\[2mm]
    \label{startsalviosanityboommid}
   \; = \; &
   \, \frac{ d^{ }_3 \kappa_{ }^2 T  }{8\pi^2}
    \int_{\mathcal{O}(g_3^2 T)}^{ }
   \frac{
     \mathrm{d}q \,q_{ }^2
   }{
     q ( 1+b^{ }_3 ) - a^{ }_3
   } 
   \arctan \frac{
     q_0^* \pi m_\rmii{D3}^2
   }{
     4q_{ }^2 \big[q(1+b^{ }_3)-a^{ }_3\big]
   }
   \\[2mm]
      \label{startsalviosanityboom}
   \; \approx \; & 
   \, \frac{ 
      d^{ }_3 \kappa_{ }^2 T
   }{
     16\pi(1+b^{ }_3)
   }
    \int^{ }_{\mathcal{O}(g_3^2 T)} 
    \frac{ 
      \mathrm{d}q\,q^2
    }{
      \left\vert q-\frac{a^{ }_3}{1+b^{ }_3}\right\vert
    }\to \infty
    \; ,
\end{align} 
where we replaced only the HTL piece from \eq\eqref{impartTsmallfreq}
in \eq\eqref{spf_T}, as it yields the leading term above. 
The cut-off on the $q^0_{ }$ integral is chosen such that
$g_3^4T\ll q_0^*\ll g_3^2 T$, 
 and thus isolates
the divergent region where these approximations are valid. 
The coefficients $a_3$ 
and $b_3$ can be extracted from \eq\eqref{trouble} as
\begin{equation}
    \label{aandb}
    a^{ }_3 
    \; = \; 
    \frac{3 g_3^2   T}{16}
    \; ,
    \qquad
    b^{ }_3 
    \; = \;  
    -\frac{ 6 g_3^2 }{(4\pi)^2} 
    \bigg[ 
      \ln\frac{ 4 \pi T }{ e^{\gamma_E} \bar\mu }
      +  \bigg( 1 - 4\ln2 \bigg )
    \bigg]
    \; \stackrel{\bar\mu=\pi T}{=} \; 
    5.78106 \times \frac{g_3^2 }{(4\pi)^2}\,.
\end{equation}
Evidently $a^{ }_3/(1+b^{ }_3)>0\,$, which is 
the reason for the divergence in \eq\eqref{startsalviosanityboom}.
It is worth remarking that, if HTL propagators are used, \eq\eqref{startsalviosanityboom}
still applies for the $q\lesssim g_3^2 T$ contribution; 
in which case $a^{ }_\text{HTL} = b^{ }_\text{HTL} = 0$ 
and no problem occurs, i.e.
\begin{equation}
  \label{linslope}
  \frac{\mathrm{d}\Gamma_\psi(k)}{\mathrm{d}q}
  \bigg\vert^\text{HTL}_\text{gauge}
  \; \stackrel{q\lesssim g_3^2 T}{=} \;
  \frac{ d^{ }_i \kappa_{ }^2 T  }{ 16\pi } q\,.
\end{equation}
As we will show later, it also applies in this regime in 
more generically divergence-free cases 
when $a^{ }_i/(1+b^{ }_i) \le 0$.

Equations~\eqref{startsalviosanityboom} and \eqref{linslope} show how
the $q\lesssim g_3^2T$ contribution is independent of $\mD$ and of the soft
scale more generally and only contains the ultra-soft one. In the HTL
case in \eq\eqref{linslope} this shows up as a quadratic sensitivity,
meaning that $\int\mathrm{d}q\,q\sim (g_3^2T)^2$. If one were to 
properly deal with this region, perhaps using the lattice 3D EFT as 
in Ref.~\cite{Moore:2019lgw} or the classical lattice gauge theory
as done in Ref.~\cite{Bouzoud:2026rur} for the axion
then the linear-in-$q$ behaviour would appear
as the UV asymptote of that region. The Feynman-gauge result 
in \eq\eqref{startsalviosanityboom},
while also formally of size $(g_3^2T)^2$, is on the other hand 
 divergent due to its gauge-dependent resummation. 

In the analysis of axion production from a SM plasma~\cite{Bouzoud:2024bom},  
this gauge-dependent pole appeared in Feynman gauge at
$q\approx 3 g_3^2 N^{ }_3 T/16=9g_3^2 T/16$. 
In the present case this has shifted to 
$q\approx a^{ }_3=g_3^2 T(3N^{ }_3-N^{ }_{s\,3})/16=3g_3^2 T/16$ 
due to the positive, gauge-invariant contribution of the soft squark loop.
Differently from the case of the axion in a SM plasma, the SU(2) contribution is in this 
case not affected by this divergence in Feynman gauge, as anticipated in footnote~\ref{foot_scalars}.
This is because $N^{ }_{s\,2} = T_F (N_g(1+3)+2) = 7>3N_2$ 
(the $N^{ }_g=3$ slepton and 9 squark doublets and the two Higgs doublets, 
with an overall $T_F=1/2$ normalisation). For completeness $N^{ }_{f\,2}=T_F[N^{ }_g(1+3)+2] + T_A=9$ 
from the left-handed lepton and quark doublets, the Higgsinos and the Winos.

In the abelian case the issue is absent, since there is no ultra-soft scale there and the 
one-loop polarisation tensor is gauge-invariant. There  $N_{s\,1}=N_{f\,1}=11$ after summing 
over the hypercharges squared. Note that in all three cases SUSY implies
equal numbers of fermions and bosons, i.e. $\ncol+\nscal=\nferm$.

%
\begin{figure}[t]
\centerline{
  \includegraphics[width=0.8\linewidth]{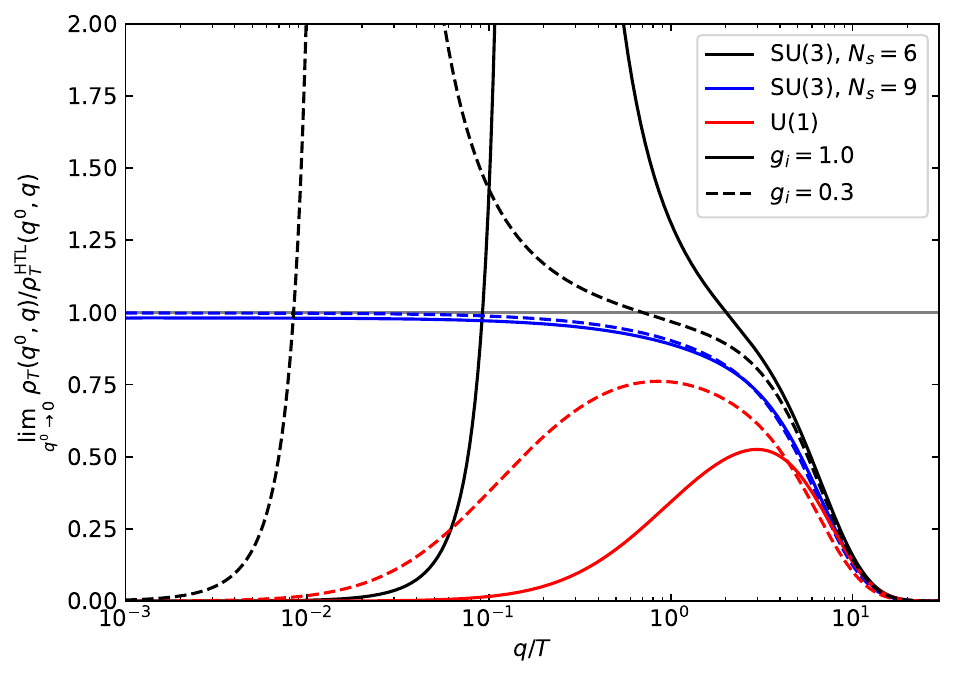}
}
\vspace{-3mm}
\caption[a]{\small
  The Feynman-gauge resummed transverse spectral function 
  from \eqref{spf_T}, as a function of $q\,$, for 
  zero frequency ($q^{0}_{ }=0$),  normalized by its HTL version.
  The black curves are for the gluon spectral function
  and the red ones are for the U(1) gauge boson spectral function, 
  both with MSSM matter content. 
  The blue curves give an adjusted gluon spectral 
  function with $N_3=3$ and $N_{f\,3}=9$ but, unlike in the MSSM, 
  $N_{s\,3}=9$, 
  thus removing the leading term in \eq\eqref{trouble}. 
  The solid and dashed curves are for $g^{ }_i =1$ and 
  $g^{ }_i = 0.3$ respectively,
  with $\bar\mu=\pi T$ in all cases.
}
\label{fig:spfs}
\end{figure}
%

To illustrate these different behaviors, in \fig\ref{fig:spfs}, 
we plot the zero-frequency
limit of the transverse spectral function normalized by 
its HTL counterpart. 
From the previous discussion
the latter reads 
$
  \lim^{ }_{q^0\to0}\;
  \rho_T^\mathrm{\small HTL}(q_{ }^0,q)
  =
  \pi  \mDi^2 q^0_{ }/(2 q_{ }^5)
\,$. 
For the MSSM SU(3) curves in black, we see that both at 
$g_3=1$ ($m_{\rmii{D}3}\approx 2.1\,T$) 
and at 
$g_3=0.3$ ($m_{\rmii{D}3}\approx 0.64\,T$) 
the gauge-dependent divergent structure appears clearly at 
$q\approx 3g_3^2 T/16\,$, 
markedly 
deviating from the HTL result, 
before being exponentially suppressed
at large $q/T$ 
stemming from 
the behavior of
$\mathrm{Im}\,\Pi_R^T(0, q\gg T)$ which 
is not captured by the HTL approximation when
extrapolated outside its validity domain.

Conversely, the U(1) curves in red show no trace of the poles, as expected, since $N_1=0$. The large 
suppression at small values of $q/T$ can be understood as originating from the 
\emph{positive} $ g_1^2  N_{s\,1} q T/16$ term---or equivalently the \emph{negative} 
$ a_1=-11g_1^2T/16$---which dominates the denominator in \eq\eqref{spf_T} for 
$q\ll g_1^2T$. 
Finally, we show in blue the SU(3) contribution in a theory where $N_{s\,3}=9$, thus exactly 
canceling---in Feynman gauge---the leading term in \eq\eqref{trouble}. In this case the resulting curves are very close to the HTL 
ones in the IR, as they can only differ by the corresponding $b_3$ term of 
order $g_3^2/(4\pi)^2$, to then give way to exponential 
suppression in the UV. This is the behavior that one would also find in abelian theories 
with no charged scalars, which are also free of the leading, linear term in $q$ in \eq\eqref{trouble}.

In deriving \eq\eqref{startsalviosanityboom} we have worked for illustration 
with bare gauginos. If they are instead kept in resummed 
form~\cite{Rychkov:2007uq,Eberl:2024pxr}, \eq\eqref{bareplasmon}
would have to be replaced by its resummed counterpart. 
As the pathological regime arises for 
$\vert q^0\vert \ll q\approx 3g_3^2 T/16$ in the SU(3) case,
for $k\gtrsim T$ we can approximate  the resummed spectral function 
from \eqref{spf_pm},
by its asymptotic regime,
which reads
\begin{equation}
  \rho^{\text{pole}}_+ \big( p \gg g_i T \big)
  \; \approx \;
  2\pi\delta \bigg( \underbrace{(k-q_{ }^0)}_{p_{ }^0} 
  -\, p - \frac{m_{\infty i}^2}{2p} 
  \bigg)
  \; ,
  \label{UVplasmon}
\end{equation}
where $m_{\infty i}^2=\mDi^2/2$ is the asymptotic mass of the gaugino. 
If we look at the $q^0=0$ region, \eq\eqref{UVplasmon} 
then implies $p\approx k-m_{\infty 3}^2/(2k)$. 
Combining this with $p+q > | \vec{p} + \vec{q} | = k\,$, as for the domains in \fig\ref{fig:poles_and_cuts},  
then yields $q>m_{\infty 3}^2/(2k)$. 
The pole at $q\approx 3g_3^2 T/16$ is thus in the 
integration region if $k\gtrsim 6 T$ and causes
a divergent contribution of the form of \eq\eqref{startsalviosanityboom}.\footnote{%
    \label{htl_gluino}
    In that case, for $k\gg m_\infty$ we can use the asymptotic form in Eq.~\eqref{UVplasmon} 
    for the spectral function. Upon expanding in $k\gg m_\infty\gg q\gg q_0$ we find
\begin{equation}
   \label{boomHTL}
  \Gamma_\psi(k)\bigg\vert^\text{\cite{Rychkov:2007uq} cp div}_{\text{SU(3) HTL }\tilde{g}}\approx 
\frac{ d_3 \kappa^2 T  }{16\pi(1+b_3)}
    \int_{\mathcal{O}(g_3^2 T)} \frac{ \mathrm{d}q}{\left\vert q-\frac{a_3}{1+b_3}\right\vert}
    \left(q^2+\frac{m_{\infty3}^4}{4k^2}\right)\theta\left(q-\frac{m_{\infty3}^2}{2k}\right)\,.
\end{equation}
}
When studying axion production, we have two differences. First, in the SM 
$m_{\infty 3}^2=m_\rmii{D3}^2/2=g_3^2 T^2(1+N_f/6)/2$,
with $N_f$ now the number of quarks obeying $T\gg m_q$. Second, the Feynman-gauge pole  occurs at $q\approx 9g_3^2 T/16$. 
Hence the pole is in the integration range for $k/T\gtrsim 4(1+N_f/6)/9\ll 6$.
This might then explain why the Feynman-gauge results of Refs.~\cite{Salvio:2013iaa,DEramo:2021lgb} for axion
production are a factor of 5 to 10 larger
than those in Ref.~\cite{Bouzoud:2024bom} coming from the tuned scheme for $g_3\gtrsim 1$, whereas in the gravitino case 
they are of comparable size. This of course assumes that the artificial regularisation mechanism is 
the same in the two cases.

%
\begin{figure}[t]
\centerline{
  \includegraphics[width=0.8\linewidth]{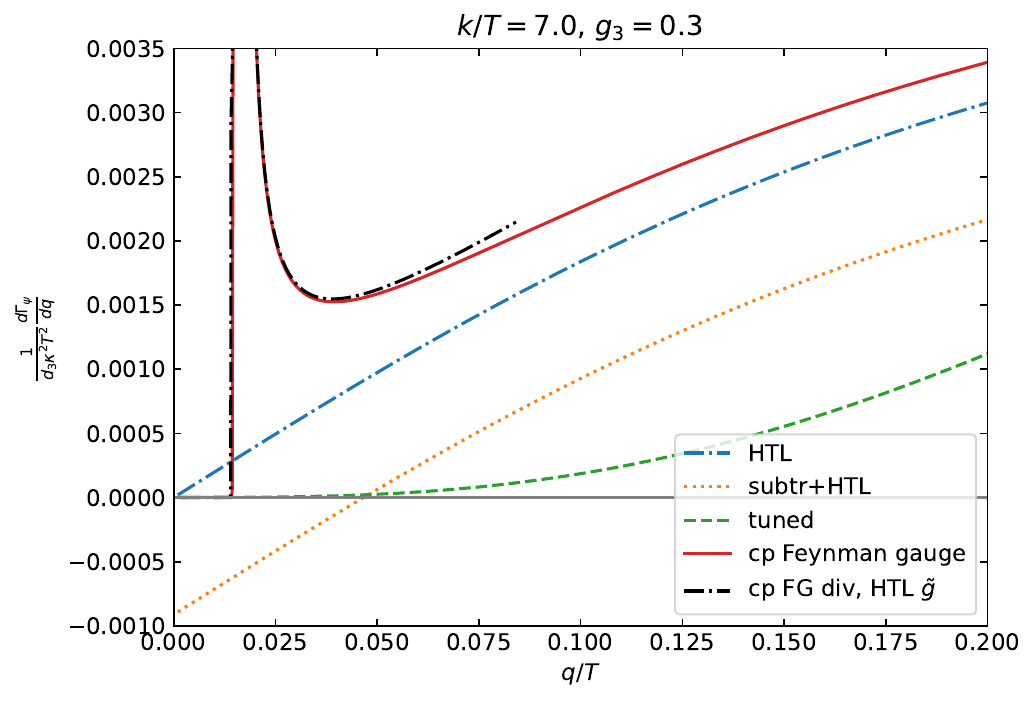}
}
\vspace{-3mm}
\caption[a]{\small 
  The ``HTL'' curve comes from replacing the Feynman-gauge spectral functions with their HTL
  counterparts in \eq\eqref{salviosanity3} and performing the $q_0$ integration numerically.
  The ``subtr+HTL'' curve adds to the ``HTL'' curve
  the $g_3^2$-proportional part of the first three lines in
  \eq\eqref{subtrlog}, kept differential in $q$. 
  The ``tuned'' curve is the differential-in-$q$ version of the $g_3^2$ contribution to
  \eq\eqref{tuned}.
  The ``cp Feynman gauge'' curve comes from 
  numerical integration of 
  the cut-pole ``+ plasmon'' 
  contribution to \eq\eqref{startsalvio2}---which amounts 
  to accounting for the HTL dispersion relation of the ``particle'' 
  mode of the gluino. Finally, the ``cp FG div, HTL $\tilde{g}$'' curve 
  is given in \eq\eqref{boomHTL}. It
  is the analytical approximation to the 
  ``cp Feynman gauge'' curve
  accounting for the dominant, divergent contribution.
}
\label{fig:div}
\end{figure}
%

It is helpful to plot a more differential version of the production rate, 
$\mathrm{d}\Gamma_\psi(k)/\mathrm{d}q$ 
as a function $q$, so that the area under the curve 
gives the production rate. 
We do this in \fig\ref{fig:div} 
for $k = 7T$, displaying the various schemes. 
First, the cp contribution 
using the HTL gaugino pole for $|p_0|> p$ in \eq\nr{spf_pm} 
and the FG spectral functions for $|q_0| < q$ in \eqs\nr{spf_T} 
and \nr{spf_L}. 
This corresponds to the prescription in Ref.~\cite{Rychkov:2007uq}. 
Secondly, the original ``soft'' computation amounts to implementing 
HTL spectral functions in \eq\nr{salviosanity3}. 
This should then be added to a ``hard'' computation 
which corresponds to the first three lines in \eq\nr{subtrlog} 
(made differential\footnote{%
\label{foot_subtr} 
Doing so, and subsequently expanding for small $q$ we 
find a constant shift:  
\begin{equation}
  \frac{\mathrm{d} \Gamma_\psi(k)}{\mathrm{d}q}
  \bigg\vert^\text{subtr+HTL}
  \; = \;  
  \frac{\mathrm{d}\Gamma_\psi(k)}{\mathrm{d}q}
  \bigg\vert^\text{\cite{Rychkov:2007uq} cp HTL}_\text{gauge}
  -\ 
  \frac{
    \left(8+\pi ^2\right) d_i g_i^2 \kappa ^2 T^2 (\ncol+\nscal)
  }{
    512\pi^3
  }
  \, + \, \mathcal{O}(q T)
  \; .
\end{equation}
} in $q$), 
and would be in the spirit of the original calculation~\cite{Bolz:2000fu}. 
Lastly, we also show a result 
corresponding to the ``tuned'' screening mass from \eq\eqref{tuned}. 

%
\begin{figure}[t]
\centerline{
  \includegraphics[width=0.9\linewidth]{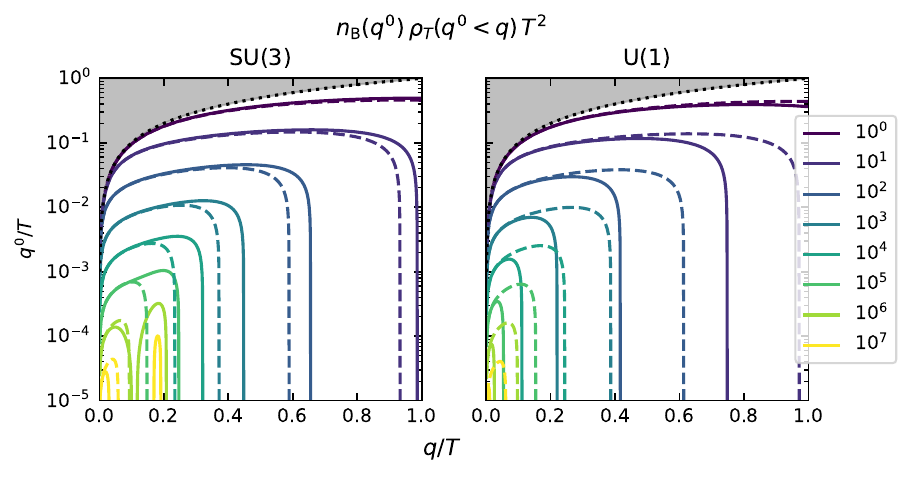}
}
\vspace{-3mm}
\caption[a]{\small 
  The space-like Feynman-gauge and HTL resummed transverse spectral 
  functions multiplied by the Bose distribution.
  Feynman-gauge contour lines are solid, HTL ones are dashed. 
  The dotted black line is the $q^0=q$ boundary separating the space-like 
  and time-like domains; the latter is shaded in this figure. 
  The gauge couplings are set to $g_i=1$ with MSSM matter content,
  $\bar\mu=\pi T$.
}
\label{fig:contours}
\end{figure}
%

Finally, we speculate on the numerical artifacts that lead to finite results in 
\cite{Rychkov:2007uq,Eberl:2020fml,Eberl:2024pxr}. 
As was argued in detail in~\cite{Bouzoud:2024bom}, the narrow, gauge-dependent divergent structure in 
non-abelian transverse gauge boson 
spectral functions---SU(3) only for MSSM processes, SU(3) and SU(2) for SM ones
relevant e.g. for axion production---can easily be missed or artificially regulated numerically, 
in particular once interpolators are used for the thermal parts of the polarisation tensor or 
for the spectral function itself. Indeed, Ref.~\cite{Salvio:2013iaa} provides contour plots
of the  Feynman-gauge and HTL resummed gauge-boson spectral function\footnote{%
It is unclear which gauge boson is being plotted there. The SU(3) and SU(2) ones should contain the divergent
structure in the SM plasma considered there.} on linear scales of frequency $q^0$ and 
momentum $q\,$. If these are used as the basis for interpolators, insufficient resolution might miss
the pole entirely, given that it lies at vanishingly small frequencies, where the spectral 
function itself vanishes identically.

The relevant quantity is however the $G^<_T$ Wightman function, 
namely $G^<_T(\Q)=\nB(q^0)\times \rho_T(q^0,q)$, which appears in \eq\eqref{startsalvio2}. 
In \fig\ref{fig:contours} we plot contours for 
these quantities for the SU(3) and U(1) gauge bosons in the MSSM on a logarithmic frequency scale. The gauge-dependent divergent structure in the SU(3) case is now clearly visible around
$q\approx 3g_3^2 T/16\approx 0.19 T$ for $g_3=1\,$.

\newpage

{\small
%
\bibliographystyle{utphys}
\bibliography{ref.bib}

\cleardoublepage
}

\end{document}